\documentclass[A4paper,twoside,journall]{IEEEtran}
\usepackage[T1]{fontenc}
\usepackage{algorithm,amsbsy,amsmath,amssymb,epsfig,bbm,mathrsfs,multirow,amsthm}
\usepackage{soul}
\usepackage{enumitem}
\usepackage{subfig}
\usepackage{enumitem}
\usepackage{array,multirow,graphicx}
\usepackage{nicefrac}
\usepackage{soul}
\usepackage{flushend}
\usepackage{epstopdf}
\usepackage{color}
\usepackage{soul}
\usepackage{bbm}
\usepackage{cite}
\usepackage{algpseudocode}
\usepackage{mathtools}
\usepackage{pdfpages}
\usepackage{booktabs}

\usepackage{subcaption}
\usepackage{amsmath,verbatim}
\usepackage{cases}

\usepackage[table]{xcolor}
\usepackage{booktabs}

\newcommand{\beq}{\begin{equation}}
\newcommand{\eeq}{\end{equation}}
\newcommand{\bitm}{\begin{itemize}}
\newcommand{\ba}{\begin{array}}
\newcommand{\ea}{\end{array}}
\newcommand{\eitm}{\end{itemize}}
\newcommand{\beqn}{\begin{eqnarray}}
\newcommand{\eeqn}{\end{eqnarray}}
\newcommand{\beqno}{\begin{eqnarray*}}
\newcommand{\eeqno}{\end{eqnarray*}}
\newcommand{\bma}{\begin{displaymath}}
\newcommand{\ema}{\end{displaymath}}
\newcommand{\bnu}{\begin{enumerate}}
\newcommand{\enu}{\end{enumerate}}
\newcommand{\bce}{\begin{center}}
\newcommand{\ece}{\end{center}}
\newcommand{\btb}{\begin{tabular}}
\newcommand{\etb}{\end{tabular}}

\usepackage{hyperref}
\hypersetup{
     unicode=false,          
     pdftoolbar=true,        
     pdfmenubar=true,        
     pdffitwindow=false,     
     pdfstartview={FitH},    
     pdftitle={My title},    
     pdfauthor={Author},     
     pdfsubject={Subject},   
     pdfcreator={Creator},   
     pdfproducer={Producer}, 
     pdfkeywords={keyword1, key2, key3}, 
     pdfnewwindow=true,      
     colorlinks=true,       
     linkcolor=red,          
     citecolor=red,        
     filecolor=red,      
     urlcolor=red           
}
\usepackage{xcolor,cite,etoolbox}
\makeatletter 
\pretocmd\@bibitem{\color{black}\csname keycolor#1\endcsname}{}{\fail}
\newcommand\citecolor[1]{\@namedef{keycolor#1}{\color{blue}}}
\makeatother

\usepackage{titlesec}
\titlespacing*{\section}{0pt}{0.4\baselineskip}{0.4\baselineskip}
\titlespacing*{\subsection}{0pt}{0.2\baselineskip}{0.2\baselineskip}

\begin{document}

\markboth{IEEE Transactions on Communications}{Hoang \MakeLowercase{\textit{et al.}}: Securing SIM-Assisted Wireless Networks via Quantum Reinforcement Learning}

\title{Securing SIM-Assisted Wireless Networks via Quantum Reinforcement Learning}

 \author{ 
 {{Le-Hung Hoang}, {Quang-Trung Luu}, Dinh Thai Hoang, Diep N. Nguyen, and Van-Dinh Nguyen}

\thanks{Part of this work was presented at the IEEE International Conference on Advanced Technologies for Communications, Oct. 2025, Hanoi, Vietnam~\cite{hoang2025secure}. This work is supported by the project VUNI.GREEN-X.RISE.AY25-27.03.}  

\thanks{L.-H. Hoang is with the Smart Green Transformation Center, VinUniversity, Vinhomes Ocean Park, Hanoi 100000, Vietnam (e-mail: hung.hl@vinuni.edu.vn).}%
\thanks{Q.-T. Luu is with Universit\'e Paris-Saclay - CNRS - CentraleSup\'elec - L2S, Gif-sur-Yvette, F-91192, France (e-mail: quangtrung.luu@centralesupelec.fr)}%
\thanks{D. T. Hoang and D. N. Nguyen are with the School of Electrical and Data Engineering, University of Technology Sydney, Sydney, NSW 2007, Australia (e-mails: \{hoang.dinh, diep.nguyen\}@uts.edu.au).}
\thanks{V.-D.~Nguyen (corresponding author) is with the School of Computer Science and Statistics, Trinity College Dublin, The University of Dublin, Dublin 2, D02 PN40, Ireland (e-mail: dinh.nguyen@tcd.ie).}%
}
\maketitle
\begin{abstract}
Stacked intelligent metasurfaces (SIMs) have recently emerged as a powerful wave-domain technology that enables multi-stage manipulation of electromagnetic signals through multilayer programmable architectures. While SIMs offer unprecedented degrees of freedom for enhancing physical-layer security, their extremely large number of meta-atoms leads to a high-dimensional and strongly coupled optimization space, making conventional design approaches inefficient and difficult to scale. Moreover, existing deep reinforcement learning (DRL) techniques suffer from slow convergence and performance degradation in dynamic wireless environments with imperfect eavesdropper channel state information (CSI). To address these challenges, we propose a hybrid quantum proximal policy optimization (Q-PPO) framework for SIM-assisted secure communications that jointly optimizes transmit power allocation and SIM phase shifts to maximize the average secrecy rate under power and quality-of-service constraints. Specifically, a parameterized quantum circuit is embedded into the actor network, forming a hybrid classical-quantum policy architecture that enhances policy representation capability and exploration efficiency in high-dimensional continuous action spaces. Extensive simulations demonstrate that the proposed Q-PPO scheme consistently outperforms DRL baselines, achieving approximately $15\%$ higher secrecy rates and $30\%$ faster convergence under imperfect eavesdropper CSI. These results establish Q-PPO as a powerful optimization paradigm for SIM-enabled secure wireless networks.
\end{abstract}


\section{Introduction}
\label{sec:introduction}
\IEEEPARstart{T}{he} sixth-generation (6G) networks are expected to revolutionize wireless communications by delivering significant improvements over 5G in terms of data throughput, energy efficiency, and massive connectivity~\cite{8766143}. However, the explosive growth of connected devices, together with the broadcast nature of wireless channels, substantially increases the risk of confidential data leakage to unauthorized receivers. As a result, emerging 6G applications demand not only ultra-low latency and high-precision localization but also stringent security guarantees, thereby imposing more rigorous requirements on information privacy protection.

To meet these security requirements, physical-layer security (PLS) has emerged as a promising complementary paradigm for safeguarding wireless communications~\cite{yang20221,jin2021,NguyenTIFS16}. Unlike conventional cryptographic approaches that rely on key distribution and encryption overhead, PLS exploits the intrinsic randomness of wireless channels to enhance confidentiality directly at the physical layer. A variety of PLS techniques have been widely investigated. For instance, artificial noise generation~\cite{goel2008} deliberately injects interference to degrade the eavesdropper’s channel while preserving the legitimate link, whereas secure beamforming (\textit{e.g.},~\cite{NguyenJSAC2018} and \cite{11020999}) steers transmission energy toward intended receivers to suppress information leakage. Despite their effectiveness, these approaches often require additional transmit power, cooperative nodes, or complex hardware architectures, and may fail to guarantee secure communication in certain network configurations.

Driven by the need for more energy-efficient and flexible security solutions, reconfigurable intelligent surfaces (RISs) have recently attracted considerable attention. By adaptively adjusting the amplitude and/or phase of incident electromagnetic (EM) waves, RISs enable passive reconfiguration of the wireless propagation environment, providing an energy-efficient and scalable complement to conventional active beamforming through additional degrees of freedom for wavefront manipulation~\cite{10298597,11268332,10534116}. However, most existing RIS-assisted secure transmission schemes rely on single-layer metasurface architectures~\cite{10298597,11268332,10534116,10266977}. This structural simplicity inherently limits the available degrees of freedom for wavefront shaping and makes such systems less effective in suppressing multiuser interference, thereby still necessitating complex digital beamforming at the transceiver side~\cite{8982186}.

To overcome these limitations, stacked intelligent metasurfaces (SIMs) have recently been proposed as a powerful alternative~\cite{10922857,papazafeiropoulos2024achievable,hassan2024efficient,liu2025stacked}. Inspired by artificial neural network architectures, SIMs consist of multiple cascaded metasurface layers that enable fine-grained, multi-stage control of EM wave propagation. By flexibly adjusting the phase shifts of meta-atoms across layers, SIMs can transform the input carrier signal into a desired output waveform, facilitating highly precise beamforming~\cite{an2025emerging,liu2025stacked}. From a PLS perspective, the multi-layer structure of SIMs provides substantially more degrees of freedom for shaping the propagation environment, allowing the transmitted wavefront to be optimized to strengthen legitimate links while deliberately degrading eavesdropping channels~\cite{10679315,10767193,kavianinia2025secrecy}. Compared with conventional single-layer RIS designs, SIMs enable multiple cascaded transformations, which significantly enhance interference suppression and reduce hardware costs in multiuser beamforming scenarios without requiring large antenna arrays, numerous radio frequency (RF) chains and high-resolution digital-to-analog converters (DAC)~\cite{papazafeiropoulos2024achievable,hassan2024efficient}. These advantages position SIMs as a promising enabler for next-generation physical-layer security solutions.

\subsection{Related Works}
Existing studies on SIM-enabled wireless communication systems have mainly focused on wave-domain beamforming optimization~\cite{10279173,10922857,10158690,papazafeiropoulos2024achievable,hassan2024efficient}. The SIM concept was first introduced in~\cite{10279173} for multi-user multiple-input single-output (MISO) downlink systems, where joint optimization of transmit power allocation and SIM phase shifts was proposed to maximize the system sum rate. The results demonstrated that SIM-assisted architectures can significantly improve the performance of conventional MISO systems while reducing precoding complexity and latency. Subsequently,~\cite{10922857} extended this framework by enabling independent phase control of individual meta-atoms, providing more flexible wave manipulation across multiple SIM layers. In~\cite{10158690}, SIMs were further deployed at both the transmitter and receiver in point-to-point multiple-input multiple-output (MIMO) systems, enabling direct signal coding and combining in the electromagnetic domain.

Beyond beamforming, SIMs have also been explored as wave-domain information processing platforms. For instance, SIM-assisted channel estimation has been investigated in~\cite{yao2024channel,an2024hybrid}. Specifically,~\cite{yao2024channel} addressed the underdetermined channel estimation problem by exploiting multiple pilot observations and proposing a reduced-subspace least-squares estimator that leverages the rank deficiency of spatial correlation matrices. Hybrid digital–wave domain estimation strategies were developed in~\cite{an2024hybrid} to minimize mean square error, achieving performance comparable to fully digital architectures even with limited RF chains. Moreover, SIMs have recently been utilized as semantic encoders in the wave domain~\cite{huang2024stacked,li2021spectrally}. The work in~\cite{huang2024stacked} demonstrated that SIM can map input images into class-dependent beam patterns for low-complexity semantic detection, while~\cite{li2021spectrally} showed that a three-layer SIM can perform high-accuracy object classification by transforming semantic features into spectral power distributions. Notably, these studies did not consider secure transmission or the presence of eavesdroppers.

More recently, SIMs have been introduced into PLS design~\cite{10679315,10767193,kavianinia2025secrecy}. In~\cite{10679315,10767193}, SIM-based anti-eavesdropping strategies were proposed for single-input single-output systems, where SIM simultaneously performs modulation, beamforming, and artificial noise generation to enhance secrecy and energy efficiency. The work in~\cite{kavianinia2025secrecy} further investigated sum secrecy-rate maximization in SIM-assisted multi-user MISO systems under transmit power constraints, and proposed an alternating optimization framework based on successive convex approximation and projected gradient ascent to address the non-convex optimization problem induced by discrete SIM phase shifts. Although these studies demonstrated the considerable potential of SIMs for secure wireless communications, they generally relied on the idealized assumption of perfect eavesdropper channel state information (CSI), which is difficult to obtain in practical wireless environments.

Furthermore, most existing SIM optimization frameworks adopt one-shot model-based optimization, resulting in highly coupled and non-convex problems that must be repeatedly solved at each time slot, leading to high computational complexity and sensitivity to initialization~\cite{an2025emerging}. To overcome these limitations, deep reinforcement learning (DRL) has recently emerged as a promising alternative for adaptive SIM control in dynamic wireless environments~\cite{10689487,11020999}. By learning policies directly through interaction with the environment, DRL-based schemes can adapt to time-varying channels without requiring complete system knowledge. For example,~\cite{10949617} employed a deep deterministic policy gradient (DDPG) algorithm for joint SIM phase shift and transmit power optimization, demonstrating superior sum-rate performance compared to conventional optimization methods~\cite{yang2024}. However, classical DRL approaches often suffer from slow convergence and scalability issues in high-dimensional SIM control problems, motivating the need for more efficient learning paradigms.

\subsection{Motivations and Main Contributions}
Despite the promising secrecy gains offered by intelligent metasurfaces, practical deployment of SIM-enhanced PLS remains severely hindered by high-dimensional control complexity. The multilayer SIM architecture introduces a massive configuration space, leading to a highly coupled and dynamic control problem for real-time phase shift adaptation in time-varying wireless environments. This dimensional explosion fundamentally challenges the scalability of existing DRL approaches. Conventional DRL methods typically rely on large-scale neural networks (NNs) to approximate high-dimensional policies, which often leads to substantial computational overhead, slow convergence, and prohibitive training complexity~\cite{hoang2023deep}. These limitations highlight the urgent need for more efficient and scalable learning frameworks for SIM-enabled secure communications.

Recent advances in quantum machine learning (QML) have introduced a new computational paradigm that leverages quantum mechanical principles to enhance learning and optimization capabilities~\cite{hassan2025quantum}. A particularly promising branch of QML is quantum reinforcement learning (QRL), which offers a fundamentally different mechanism for addressing high-dimensional control problems. By exploiting quantum superposition and entanglement, QRL enables the simultaneous representation and evaluation of multiple state–action pairs within compact quantum states, thereby significantly enhancing exploration efficiency and accelerating convergence~\cite{liu2024qtrl}. Moreover, quantum entanglement preserves intrinsic correlations across the state–action space, allowing learning updates in one region to propagate globally and reducing the number of samples required to approach optimal policies~\cite{cherukuri2025q,chen2024introduction}. These properties make QRL particularly well-suited for large-scale SIM configuration tasks.

Motivated by these observations, this paper proposes the first hybrid QRL framework for SIM-assisted PLS under imperfect eavesdropper CSI. We investigate a SIM-enhanced multi-user MISO downlink system and develop a quantum proximal policy optimization (Q-PPO) algorithm that jointly optimizes transmit power allocation and SIM phase shifts to maximize the long-term average secrecy rate (ASR). The main contributions of this paper are summarized as follows:
\begin{itemize}
\item We investigate a SIM-assisted multi-user MISO downlink system in the presence of an untrusted user (\textit{i.e.}, an eavesdropper (Eve)) with imperfect CSI. Unlike most existing SIM-assisted security studies that assume perfect knowledge of the eavesdropper’s channel, we explicitly model CSI uncertainty and formulate a long-term ASR maximization problem through the joint optimization of transmit power allocation and SIM phase shift configurations under transmit power and quality-of-service (QoS) constraints. This formulation provides a more realistic and practically relevant framework for SIM-enabled secure wireless communications.
    
\item We propose a hybrid quantum-classical PPO algorithm, termed Q-PPO, to address the high-dimensional and strongly coupled SIM optimization problem. By embedding parameterized quantum circuits (PQCs) into the actor network~\cite{QPPO}, the proposed framework leverages quantum-enhanced feature representation and exploration mechanisms for continuous control. The hybrid architecture improves exploration efficiency and convergence behavior while reducing the dependence on large-scale classical neural networks for learning SIM phase shift and transmit power allocation policies.
      
\item Through extensive simulations, we demonstrate that Q-PPO consistently outperforms state-of-the-art (SOTA) DRL methods such as PPO, DDPG, and twin delayed DDPG (TD3) as well as existing SIM approaches~\cite{10949617,yang2024} in terms of ASR, convergence speed, and robustness to eavesdropper CSI uncertainty. We further provide quantitative insights into how the number of qubits and the depth of the quantum circuit affect learning efficiency and secrecy performance, and show that SIM architectures themselves offer substantial secrecy gains under dynamic and uncertain wireless conditions.
\end{itemize}

The remainder of this paper is organized as follows. Section \ref{sec:system_model} describes the system model, formulates the secrecy-rate maximization problem, and reviews the Markov decision process (MDP) framework used to characterize the system dynamics and uncertainties. Section \ref{Sec:QRL} presents the fundamentals of reinforcement learning and details the proposed Q-PPO algorithm. Section \ref{Sec:simulation} reports numerical results and analyzes the performance under various scenarios. Finally, Section \ref{sec:conclusion} concludes the paper.

\noindent\textit{Notation:} Throughout this paper, bold lowercase letters and bold uppercase letters are used to denote vectors and matrices, respectively. For a given vector $\mathbf{a}$, $\mathrm{diag}(\mathbf{a})$ denotes the diagonal matrix formed by placing the entries of $\mathbf{a}$ along its main diagonal. The superscripts $(\cdot)^{\top}$ and $(\cdot)^H$ stand for the transpose and conjugate transpose operations, respectively. The notations $|\cdot|$, $\|\cdot\|$, and $\mathbb{E}[\cdot]$ are used to represent the absolute value of a scalar, the Euclidean norm of a vector, and the statistical expectation, respectively.

\section{System Model and Problem Formulation}
\label{sec:system_model}
\begin{figure}[t]
\centering\includegraphics[width=0.95\columnwidth]{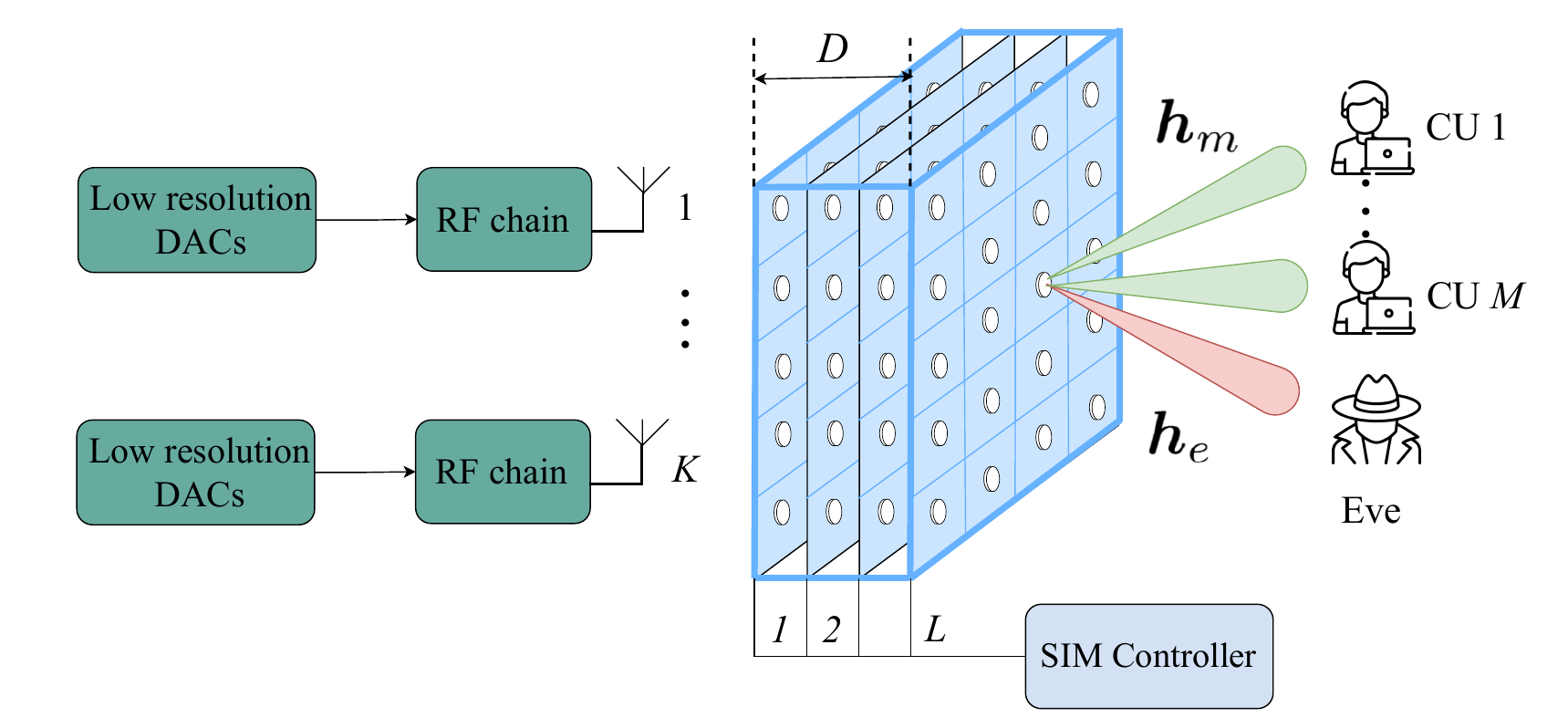}
    \caption{The SIM-aided secure multi-user communication system model.} 
    \label{fig:systemmodel}
\end{figure}

As illustrated in Fig.~\ref{fig:systemmodel}, we consider a SIM-assisted multiuser MISO downlink, in which a BS with $K$ transmit antennas serves $M$ single-antenna CUs, indexed by $\mathcal{M}\triangleq{1,\ldots,M}$. We assume $K=M$, with antenna $k=m$ carrying the data stream intended for CU $m$ with allocated power $p_m$~\cite{10922857, papazafeiropoulos2024achievable}. Under this one-to-one mapping, the SIM performs multiuser beamforming in the wave domain without requiring an additional digital precoder, allowing us to focus on joint power allocation and SIM phase-shift optimization. In addition, an untrusted user (Eve) attempts to intercept confidential messages intended for the CUs~\cite{9293148}.

\subsection{SIM Model}
The SIM is modeled as a closed vacuum container composed of multiple stacked metasurface layers~\cite{Liu2022}. Each metasurface layer contains a large number of meta-atoms~\cite{Cui2014}, which are connected to a smart controller, \textit{e.g.}, a field-programmable gate array board. By appropriately configuring the complex-valued transmission coefficients of the meta-atoms in each layer, their EM responses, specifically amplitude and phase shift, can be independently adjusted. Through this configuration, the SIM is capable of performing downlink beamforming directly in the wave domain~\cite{10279173}.

Let $\mathcal{L} \triangleq \{1, \dots, L\}$ denote the set of metasurface layers, and $\mathcal{N}_l \triangleq \{1, \dots, N\}$ represent the set of meta-atoms within layer $l \in \mathcal{L} $. Note that, in this paper, we consider square metasurface arrays, \textit{i.e.}, $N=n_{\max}^2,$ with $n_{\max} \in \mathbb{N}$. We define $\theta^l_n$ as the phase shift of meta-atom $n$ on layer $l$, which can be continuously adjusted in the range $[0, 2\pi)$. Accordingly, the diagonal phase shift matrix of the metasurface layer $l$ is defined as:
\begin{equation}
\boldsymbol{\Phi}^l=\mathrm{diag}\left(e^{j\theta^l_1},\dots,e^{j\theta^l_N}\right)\in \mathbb{C}^{N\times N}, \quad\forall l\in \mathcal{L}
\end{equation}
where $e^{j\theta^l_n}$ denotes  the transmission coefficient imposed by  meta-atom $n$ on metasurface layer $l$. Similar to~\cite{10158690}, we consider metasurfaces as uniform planar arrays. Specifically, as shown in Fig.~\ref{fig:systemmodel}, the metasurfaces rely on an isomorphic lattice arrangement and are parallel to the $xz$-plane. The distance between two different meta-atoms ($n$ and $n'$) on the metasurface can be expressed as
\begin{equation}
    d_{n,n'}= d_{\mathrm{a}}\sqrt{(n_x-{n'}_x)^2+(n_z-{n'}_z)^2}
\end{equation}
where $d_{\mathrm{a}}$ denotes the distance between adjacent meta-atoms on a metasurface. Additionally, $n_x$ and $n_z$ denote the indices of meta-atom $n$ along the $x$-axis and the $z$-axis, respectively,
which are defined as
\begin{equation}
    n_x=\mathrm{mod}(n-1,n_{\max})+1; \quad n_z = \lceil n/ n_{\max}\rceil.
\end{equation}
We denote $D$ as the distance from the antenna array to the output metasurface layer. For simplicity, we assume that the distance between adjacent metasurface layers and the distance between the antenna array and the input metasurface layer are uniform, given by $d = D / L$. Consequently, the propagation distance from meta-atom $n$ on  layer $(l - 1)$ to the corresponding meta-atom $n'$ on  layer $l$ is given by
\begin{equation}
    d^l_{n,n'}=\sqrt{d_{n,n'}^2+d^2}, \quad\forall l\in\mathcal{L}\setminus\{1\}.
\end{equation}

To address the transmission process between transmit antennas and metasurfaces, we define $\boldsymbol{W}^l\in \mathbb{C}^{N\times N}$, $l\in \mathcal{L}\setminus\{1\}$ as the transmission coefficient matrix between layers $(l-1)$ and $l$. The antenna-to-SIM propagation matrix is $\boldsymbol{W}^1\in\mathbb{C}^{N\times K}$, with entry $w^1_{n,k}$ representing propagation from antenna $k$ to meta-atom $n$ on the first layer. Under $K=M$, the column $\boldsymbol{w}^1_m\in\mathbb{C}^{N\times 1}$ contains the coefficients associated with antenna $m$, which carries the stream intended for CU $m$. According to the Rayleigh-Sommerfeld diffraction theory~\cite{aat8084}, the $(n,n')$-th entry of $\boldsymbol{W}^l$ is given by\footnote{For analytical tractability, this work adopts the commonly used continuous phase-control SIM model. Practical hardware effects, including finite-resolution phase control, insertion loss, phase noise, and inter-element coupling, may influence system performance and will be investigated in future extensions of the proposed framework.}
\begin{equation}
\label{entry}
w_{n,n'}^l {=} \frac{A_td}{(d^l_{n,n'})^2}\left(\frac{1}{2\pi d^l_{n,n'}} {-} j\frac{1}{\lambda}\right)e^{j2\pi d^l_{n,n'}/\lambda}, l\in\mathcal{L} {\setminus} \{1\}
\end{equation}
where $A_t$ is the area of each meta-atom in the SIM, and $\lambda$ is the wavelength. Similarly, the $(n,k)$-th entry $w_{n,k}^1$ of $\boldsymbol{W^1}$ is obtained by replacing $d^l_{n,n'}$ in (\ref{entry}) with $d^1_{n,k}$, respectively, where the distance from antenna element $k$ to meta-atom $n$ at the first metasurface is given by~\cite{10158690}:
\begin{align}
d^1_{n,k} = \Big[ 
&\Big( \Big(n_z - \frac{n_{\max}+1}{2} \Big) d_{\mathrm{a}} 
   - \frac{\lambda}{2} \Big( k - \frac{K+1}{2} \Big)  \Big)^2 \notag \\
&+ \Big( n_x - \frac{n_{\max}+1}{2} \Big)^2 d_{\mathrm{a}}^2 
 + d^2 \Big]^{1/2}. \label{dkn}
\end{align}
The transfer function $\boldsymbol{G}$ of the SIM is then given by:
\begin{equation}
\boldsymbol{G}=\boldsymbol{\Phi}^L\boldsymbol{W}^L\boldsymbol{\Phi}^{L-1}\dots\boldsymbol{\Phi}^2\boldsymbol{W}^2\boldsymbol{\Phi}^{1}\in\mathbb{C}^{N\times N}.
\end{equation}

\subsection{Channel Model}
For the wireless channels, we consider a quasi-static flat-fading model. Let $\boldsymbol{h}_m$ and $\boldsymbol{h}_e\in \mathbb{C}^{N\times 1}$ denote the channels between the output metasurface and CU $m$, and between output metasurface and Eve, respectively. The channel spanning from the output metasurface to CU $m$/Eve, which is modeled according to a Rician fading channel model, can be expressed as
\begin{equation}
\label{CUchan}
    \boldsymbol{h}_{\varepsilon} = \sqrt{\frac{\beta_{\varepsilon}}{1+\kappa}}\left(\sqrt{\kappa}\boldsymbol{h}_{\varepsilon}^{\mathrm{LoS}}+\boldsymbol{h}_{\varepsilon}^{\mathrm{NLoS}}\right)\in \mathbb{C}^{N\times 1}
\end{equation}
where ${\varepsilon}\in \{m,e\}$ and $\kappa$ is the Rician factor; $\beta_{\varepsilon}= C_0d_{\varepsilon}^{-\alpha}$ is the distance-dependent path loss, with $C_0$ and $d_{\varepsilon}$ being the path loss at a reference distance of 1 meter and the distance from the output metasurface to CU/Eve $m$, respectively, and $\alpha$ being the path loss exponent; $\boldsymbol{h}_{\varepsilon}^{\mathrm{LoS}}$ and $\boldsymbol{h}_{\varepsilon}^{\mathrm{NLoS}}\in \mathbb{C}^{N\times 1}$ denote the line-of-sight (LoS) and non-LoS (NLoS) components of the channel from the output metasurface to CU/Eve $m$, respectively. Similar to~\cite{10949617} and~\cite{10892229}, we adopt a spatially correlated channel model to represent the NLoS component of the channel $\boldsymbol{h}_{\varepsilon}$, \textit{i.e.}, $\boldsymbol{h}_{\varepsilon}^{\mathrm{NLoS}}\sim\mathcal{CN}(0,\boldsymbol{R})$ where $\boldsymbol{R}\in \mathbb{C}^{N\times N}$  is the spatial correlation among the channels associated with different meta-atoms on the last layer.  By considering far-field wave propagation in an isotropic scattering wireless environment, the $(n, n')$-th entry of $\boldsymbol{R}$ can be expressed as
\begin{equation}
    \boldsymbol{R}_{n,n'}=\mathrm{sinc}\left(2d_{n,n'} / \lambda \right)
\end{equation}
where $\mathrm{sinc}(\tau)=\sin(\pi\tau)/(\pi\tau)$ is the normalized sinc function. 

Let \( \psi_{\varepsilon}^a \in [0, 2\pi) \) and \( \psi_{\varepsilon}^e \in [0, \pi/2] \) represent the physical azimuth and elevation angles of CU/Eve $m$, respectively. The LoS component $\boldsymbol{h}_{\varepsilon}^{\mathrm{LoS}}$ of channel $\boldsymbol{h}_{\varepsilon}$ can be then modelled as~\cite{10892229}
\begin{equation}
    \boldsymbol{h}_{\varepsilon}^{\mathrm{LoS}} = \boldsymbol{a}_N(\psi_{\varepsilon}^a,\psi_{\varepsilon}^e)
\end{equation}
wherein the $n$-th entry is given by
\begin{equation}
\resizebox{0.95\hsize}{!}{$[\boldsymbol{a}_N(\psi_{\varepsilon}^a,\psi_{\varepsilon}^e)]_n = 
 \exp\left\{ \frac{j2\pi d_{\mathrm{a}}}{\lambda} \Big( n_x \sin{\psi_{\varepsilon}^a} \sin{\psi_{\varepsilon}^e} \notag +\, n_z \cos{\psi_{\varepsilon}^e} \Big) \right\}.$}
\end{equation}

To design an effective beamforming scheme and allocate system resources efficiently, accurate and complete CSI is essential. For legitimate CUs, CSI can typically be obtained using standard channel estimation algorithms. However, due to Eve's tendency to conceal its presence, the CSI for this entity is often uncertain. In this paper, we assume that the channel between the output metasurface and Eve is imperfect, which can be modeled as~\cite{9374975,9293148}:
\begin{equation}
\hat{\boldsymbol{h}}_{e}= \boldsymbol{h}_{e} - \Delta\boldsymbol{h}_{e}
\label{eq:imperfect_CSI}
\end{equation}
where $\boldsymbol{h}_{e}$ denotes the perfect Eve's CSI, and the CSI error vector is assumed to follow the circularly symmetric complex Gaussian distribution, \textit{i.e.}, $\Delta\boldsymbol{h}_e\sim \mathcal{CN}(\boldsymbol{0},\delta^2_{e}\boldsymbol{I})$ with $\delta_{e}^2$ being the error variance.

{Let $\boldsymbol{s} \triangleq [s_1, \dots, s_{M}]^{\top} \in \mathbb{C}^{M}$ be the transmit signal vector, with $s_m$ being the symbol information for CU $m$ with $\mathbb{E}[\boldsymbol{s}\boldsymbol{s}^{\mathrm{H}}] = \boldsymbol{I}$. We denote by $\boldsymbol{p} \triangleq[p_1,\dots,p_M]^{\top}\in \mathbb{R}^{M}$ the transmit power vector, where $p_m$ is the power allocation for CU $m$. Under $K=M$, the BS transmit vector is $\mathbf{x}
    = \operatorname{diag}\!\left(
        \sqrt{p_1},\ldots,\sqrt{p_M}
      \right)\mathbf{s}
    \in \mathbb{C}^{M\times 1}.$} Accordingly, the received signal at CU $m$ is expressed as
\begin{align}
y_m =  &\underbrace{\boldsymbol{h}_m^H\boldsymbol{G}\boldsymbol{w}^1_m\sqrt{p_m} s_m}_{\text{Desired Signal}}  \notag\\
& +  \underbrace{\sum\nolimits_{j=1,j\neq m}^{M} \boldsymbol{h}_m^H\boldsymbol{G}\boldsymbol{w}^1_j\sqrt{p_j} s_j}_{\text{Inter-user Interference}}+ n_m,\, \forall m\in \mathcal{M}
\end{align}
where $n_m \sim \mathcal{CN}(0,\sigma_m^2)$ is the additive white Gaussian noise (AWGN) with mean zero and variance $\sigma_m^{2}$. The communication signal-to-interference-plus-noise ratio (SINR) at CU $m$ then can be expressed as:
\begin{equation}
\gamma_m=\frac{|\boldsymbol{h}_m^H\boldsymbol{G}\boldsymbol{w}^1_m|^2p_m}{\sum_{j=1,j\neq m}^{M} |\boldsymbol{h}_m^H\boldsymbol{G}\boldsymbol{w}^1_j|^2p_j+\sigma_m^2}, \quad\forall m\in \mathcal{M}.
\end{equation}

Due to the uncertainty in Eve's CSI, the received signal at Eve can be expressed as follows:
\begin{align}
y_{e}=& (\hat{\boldsymbol{h}}^H_{e}+\Delta\boldsymbol{h}_{e}^H)\boldsymbol{G}\boldsymbol{w}^1_m\sqrt{p_m} s_m \notag \\
& + \sum\nolimits_{j=1,j\neq m}^{M} (\hat{\boldsymbol{h}}^H_{e}+\Delta\boldsymbol{h}_{e}^H)\boldsymbol{G}\boldsymbol{w}^1_j\sqrt{p_j} s_j+ n_{e}
\end{align}
where $n_{e} \sim \mathcal{CN}(0,\sigma_{e}^2)$ is the AWGN with mean zero and variance $\sigma_{e}^{2}$. Under the assumption of imperfect Eve's CSI, the CSI error component $\Delta \boldsymbol{h}_e$ is treated as a source of interference. Accordingly, the achievable SINR for the signal $s_m$ at Eve is given by
\begin{equation}
\hat{\gamma}_{e}^m=\frac{|\hat{\boldsymbol{h}}_{e}^H\boldsymbol{G}\boldsymbol{w}^1_m|^2p_m}{\Psi_m}
\end{equation}
where $\Psi_m \triangleq|\Delta\boldsymbol{h}_{e}^H\boldsymbol{G}\boldsymbol{w}^1_m|^2p_m+\sum_{j=1,j\neq m}^{M} |(\hat{\boldsymbol{h}}^H_{e}+\Delta\boldsymbol{h}_{e}^H)\boldsymbol{G}\boldsymbol{w}^1_j|^2p_j+\sigma_{e}^2$.

To evaluate the security performance of the system, we adopt the achievable secrecy rate as the primary performance metric, defined as:
\begin{equation}
    R_m(\boldsymbol{p}, \boldsymbol{\theta}) = \Big[\log_2(1+\gamma_m)-\log_2(1+\hat{\gamma}_{e}^m)\Big]^+
\end{equation}
where $[x]^+ = \max(x,0)$.

\subsection{Problem Formulation}
Finally, we formulate the joint optimization of SIM-based wave-domain beamforming and BS transmit power allocation as an average secrecy-rate maximization problem. Specifically, the objective is to maximize the ASR across all users, denoted by $\bar{R}(\boldsymbol{p}, \boldsymbol{\theta})$, which is expressed as
\begin{subequations}\label{optt}
\begin{alignat}{2}
  (\mathcal{P}_1):\,  \max_{
        \boldsymbol{p},
        \boldsymbol{\theta}
    } \quad & \bar{R}(\boldsymbol{p},
        \boldsymbol{\theta}) \triangleq \frac{1}{M}\sum\nolimits_{m=1}^M R_m(\boldsymbol{p}, \boldsymbol{\theta}) \label{prob:P1}\\
    \label{cons1}
        \mathrm{s.t.} \quad & \sum\nolimits_{m=1}^{M}p_m\leq P_0&\\
        \label{cons2}
    & \log_2(1+\gamma_m)\geq R_{\min},\quad\forall m\in \mathcal{M}  \\
    \label{cons3}
    & \theta^l_n \in [0,2\pi),\quad\forall n\in \mathcal{N}, l\in \mathcal{L}
\end{alignat}
\end{subequations}
where $\boldsymbol{\theta}\triangleq \big[(\boldsymbol{\theta}^1)^{\top}, (\boldsymbol{\theta}^2)^{\top},\dots, (\boldsymbol{\theta}^L)^{\top}\big]^{\top}$ with $\boldsymbol{\theta}^l\triangleq[\theta^l_1,\theta^l_2,\dots,\theta^l_N]^{\top}$, $R_{\min}$ is  the required minimum data rate and $P_0$ is the transmit power budget at BS.
Here,  \eqref{cons1} ensures that the total allocated power does not exceed the power budget $P_0$; \eqref{cons2} guarantees a minimum data rate requirement of $R_{\min}$ for all CUs $m \in \mathcal{M}$; and \eqref{cons3} enforces that each phase shift $\theta^l_n$ takes a continuous value within the range $[0,2\pi)$.

Note that due to the strong coupling between the transmit power vector  $\boldsymbol{p}$ and the SIM phase shift vector $\boldsymbol{\theta}$, as well as the continuous phase constraints on each meta-atom, problem~$(\mathcal{P}_1)$ is inherently non-convex. In addition, the large number of meta-atoms in the SIM architecture results in an extremely high-dimensional design space, which makes the practical solution of problem~$(\mathcal{P}_1)$ particularly challenging. Conventional one-shot optimization methods must be repeatedly executed at every time slot, leading to prohibitively high computational complexity and substantial processing overhead. 

Although DRL provides a promising alternative, classical DRL approaches often suffer from poor scalability and slow convergence when applied to such high-dimensional and strongly coupled SIM configuration problems. To address these challenges, this paper proposes a novel QRL algorithm that efficiently approximates the optimal policy with significantly faster convergence than conventional DRL methods. Before presenting the proposed quantum learning framework, we reformulate the original deterministic optimization problem into a stochastic optimization problem to capture the time-varying and uncertain nature of the wireless environment.

\subsection{Stochastic Optimization Problem Formulation}
\label{sec:formulation}

In practical wireless environments, channel conditions and the eavesdropper's CSI vary over time and are inherently uncertain. Solving problem~$(\mathcal{P}_1)$ in a deterministic one-shot manner would require repeated re-optimization under complete channel knowledge, which is computationally prohibitive and unrealistic. Therefore, we reformulate the problem as a stochastic sequential decision-making problem to capture environmental dynamics and enable adaptive long-term secrecy optimization.

The joint SIM configuration and power allocation problem is formulated as a sequential decision-making process and modeled within a reinforcement learning framework, represented by the tuple~$\langle{\mathcal{S}}, {\mathcal{A}}$, ${\mathcal{R}}\rangle$, where $\mathcal{S}$, ${\mathcal{A}}$, and ${\mathcal{R}}$ denote the state space, action space, and reward function, respectively.

\noindent \textit{\textbf{State space}:} The state space characterizes the observable environmental information that evolves over time. In the considered system, the CSI of legitimate users can be accurately obtained at the end of each time slot. However, as discussed in Section \ref{sec:system_model}, acquiring perfect CSI of the eavesdropper is generally infeasible. Accordingly, the state observed at time slot $t$ is defined as
\begin{equation}
\begin{split}
    \boldsymbol{s}_t=&\{\boldsymbol{h}_{m}^t, \hat{\boldsymbol{h}}_{e}^t\}_{m\in\mathcal{M}} 
\end{split}
\label{s-spc}
\end{equation}
where $\boldsymbol{h}_{m}^t$ is the perfect CSI of CU $m$ and $\hat{\boldsymbol{h}}_{e}^t$ denotes the imperfect CSI of eavesdropper.

\vspace{0.1cm}
\noindent \textit{\textbf{Action space}:} The action space consists of the control decisions selected by the agent at each time slot, including transmit power allocation and SIM phase shift configuration, and is defined as
\begin{align}
\boldsymbol{a}_t = &\{\boldsymbol{p}_t, \boldsymbol{\theta}_t\}. 
\end{align}
These actions directly determine the SIM wave-domain beamforming behavior and the BS resource allocation strategy. The agent’s decisions must satisfy the total transmit power constraint \eqref{cons1} and the SIM phase shift constraint \eqref{cons3}. Since meta-atoms operate independently, constraint \eqref{cons3} is enforced by restricting $\boldsymbol{\theta}_t$ within the action space.
Since the considered objective is ASR maximization under a fixed transmit power budget, the available transmit power is assumed to be fully utilized. Accordingly, to satisfy constraint~\eqref{cons1}, the transmit power vector $\boldsymbol{p}_t$ is normalized such that $\sum_{m=1}^{M} p_m = P_0$, following~\cite{huang2021multi}. This normalization guarantees the feasibility of every generated action and simplifies the reinforcement learning optimization by allowing the agent to focus solely on learning the optimal power allocation among users.

\vspace{0.1cm}
\noindent \textit{\textbf{Reward}:}
Following problem~$(\mathcal{P}_1)$, the agent is designed to maximize the long-term ASR while satisfying the transmit power and QoS constraints~(\ref{cons1})–(\ref{cons3}). Since the action design and output normalization explicitly enforce constraints~\eqref{cons1} and~\eqref{cons3}, reward design only needs to consider constraint~\eqref{cons2}. Consequently, the instantaneous reward at time slot $t$ is defined as
\begin{equation}
\begin{aligned}
    r_t = \begin{cases}
        \bar{R}(\boldsymbol{p}, \boldsymbol{\theta}), & \text{if } \eqref{cons2} \text{ is satisfied},\\
        0, & \text{otherwise}
    \end{cases}
    \label{reward}
\end{aligned}
\end{equation}
which encourages the agent to learn policies that achieve high secrecy performance while maintaining feasible system operation. Note that during training, the CSI error vector is independently sampled by the environment according to~\eqref{eq:imperfect_CSI} at every time step. The sampled CSI error is used internally to compute the eavesdropper SINR and the corresponding secrecy-rate reward, while remaining unobservable to the agent.

\section{Quantum Reinforcement Learning-based Solution}
\label{Sec:QRL}
\subsection{Preliminaries of Reinforcement Learning}
To address the stochastic optimization problem formulated in Section \ref{sec:formulation}, a variety of classical DRL algorithms, including DDPG~\cite{DDPG}, TD3~\cite{fujimoto2018}, and PPO~\cite{ppo}, can be employed. Among these methods, PPO has recently attracted significant attention due to its favorable scalability, stable learning behavior, and robustness in high-dimensional continuous control tasks. PPO adopts an actor-critic architecture, in which the actor network generates actions according to a stochastic policy \( \pi_{\theta_a}(\boldsymbol{a}_t|\boldsymbol{s}_t) \) based on the observed state \( \boldsymbol{s}_t \), while the critic network evaluates the expected long-term performance of the current policy and provides gradient-based feedback to guide policy updates. A key feature that distinguishes PPO from other policy-gradient algorithms is its clipped surrogate objective function, which explicitly constrains excessive deviations between successive policy updates. This mechanism effectively balances policy improvement and training stability, thereby mitigating destructive updates that may otherwise arise in high-dimensional optimization problems.

{In DRL, the agent is trained to discover policies $\pi$ that maximize the expected cumulative discounted reward over a long-term horizon:
\begin{equation}
    \max_\pi\quad
\mathbb{E}_{\pi}
\left[
\sum\nolimits_{t=0}^{\infty} \gamma^t r_{t}
\right]
\end{equation}
where \( \gamma\in[0,\,1) \) denotes the discount factor that controls the relative importance of future rewards, and \( r_{t} \) represents the instantaneous reward defined in \eqref{reward}. Within the PPO framework, the critic network estimates the state-value function \( V^\pi(\boldsymbol{s}_t) \), which quantifies the expected discounted cumulative reward achievable from state $\boldsymbol{s}_t$ under policy $\pi$. It is mathematically expressed as
\begin{equation}
    V^\pi(\boldsymbol{s}_t) = \mathbb{E}_\pi\!\Big[\sum\nolimits_{i=0}^{\infty} \gamma^i r_{t+i}\,\Big|\,\boldsymbol{s}_t\Big].
    \label{discrew}
\end{equation}}

To further reduce the variance of policy-gradient estimates while maintaining low bias, PPO incorporates the generalized advantage estimation (GAE) technique to approximate the advantage function. The GAE is given by
\begin{equation}
    \hat{A}_t = \sum\nolimits_{i=0}^\infty (\gamma\omega)^i\Big(r_{t+i} + \gamma V^\pi(s_{t+i+1}) - V^\pi(s_{t+i})\Big)
    \label{advfunc}
\end{equation}
where $\omega\in[0,\,1]$ is the GAE smoothing coefficient that governs the bias-variance tradeoff in advantage estimation by weighting multi-step temporal-difference errors over different horizons.

To maintain training stability and prevent excessively large policy updates, PPO employs a clipping mechanism that explicitly constrains the deviation between the updated policy and the previous policy. While the learning rate uniformly scales the size of updates, the clipping operation enforces a trust region around the old policy, ensuring that policy updates remain within a controlled range. Specifically, the probability ratio between the current and previous policies is restricted within the interval \( [1 - \epsilon, 1 + \epsilon] \), which can be expressed as
\resizebox{\linewidth}{!}{\parbox{1.15\linewidth}{%
\begin{subnumcases}{\texttt{clip}(\psi, 1- \epsilon, 1+ \epsilon) =}
    1 + \epsilon, & \text{if } $\psi > 1+ \epsilon$ \\
    \psi,        & \text{if } $1 - \epsilon \leq \psi \leq 1+ \epsilon$ \\
    1 - \epsilon,& \text{if } $\psi < 1 - \epsilon$
\end{subnumcases}%
}}
where $\epsilon$ is the limit adjustment parameter. Based on this operator, the clipped surrogate objective is given by
\begin{equation}
\label{lossactor}
\resizebox{0.88\hsize}{!}{
$\mathcal{L}^{\text{CLIP}}(\boldsymbol{\theta}_{\text{a}}) = \mathbb{E}_t\Bigl[\min\Bigl(p_t(\boldsymbol{\theta}_{\text{a}})\hat{A}_t, \texttt{clip}(p_t(\boldsymbol{\theta}_{\text{a}}), 1 - \epsilon, 1 + \epsilon)\hat{A}_t\Bigl)\Bigl]$}
\end{equation}
where $p_t(\theta_a) = \pi_{\theta_a^{\rm{new}}}(\boldsymbol{a}_t|\boldsymbol{s}_t)/\pi_{\theta_a^{\rm{old}}}(\boldsymbol{a}_t|\boldsymbol{s}_t)$ is the likelihood ratio between the new and old policies. Subsequently, the overall objective function is formulated as a combination of the clipped surrogate objective, the value function loss, and an entropy regularization term to encourage sufficient exploration, following the standard PPO formulation in~\cite{ppo}, and is given by
\begin{equation}
    \mathcal{L} = \mathcal{L}^{\text{CLIP}}(\boldsymbol{\theta}_{\text{a}}) + c_1 K(\pi_{\boldsymbol{\theta}_{\text{a}}}) - c_2 \mathcal{L}^{\text{VF}}(\boldsymbol{\theta}_{\text{c}})
    \label{overactorloss}
\end{equation}
where $c_1$ and $c_2$ denote the entropy and value loss coefficients, respectively. Here, $K(\pi_{\boldsymbol{\theta}_{\text{a}}})= \mathbb{E}_t \Big[ - \sum_a \pi_{\boldsymbol{\theta}}(a|s)\log\pi_{\boldsymbol{\theta}}(a|s) \Big]$ represents the entropy bonus and $\mathcal{L}^{\text{VF}}(\boldsymbol{\theta}_{\text{c}})=\mathbb{E}_t \left[ (V_{\boldsymbol{\theta}_{\text{c}}}(\boldsymbol{s}_t)- R_t)^2 \right]$ denotes the value-function loss. The policy parameters are updated by maximizing the overall objective $\mathcal{L}$ via gradient ascent. In addition, the critic network is trained by minimizing the squared value-function loss, given by
\begin{equation}
    \mathcal{L}^{\text{VF}}(\boldsymbol{\theta}_{\text{c}})=\mathbb{E}_t \left[ (V_{\boldsymbol{\theta}_{\text{c}}}(\boldsymbol{s}_t)- R_t)^2 \right]
    \label{losscritic}
\end{equation}
where $R_t=\hat{A}_t + V_{\boldsymbol{\theta}_{\text{old}}}(\boldsymbol{s}_t)$ is the target value.

Although PPO exhibits strong performance in many complex control tasks, its conservative update mechanism can inadvertently limit exploration of substantially different strategies, which may slow convergence in extremely high-dimensional optimization problems. This limitation becomes particularly pronounced in the SIM configuration in \eqref{optt}, where the action space grows rapidly with the number of meta-atoms and layers, and the decision variables are strongly coupled. Consequently, classical PPO often requires a large number of training episodes to sufficiently explore the environment. To address this scalability bottleneck, we propose a novel quantum reinforcement learning algorithm, termed Q-PPO, which is designed to approximate the optimal policy with significantly faster convergence than conventional DRL methods.

\subsection{Proposed Q-PPO Algorithm}
\label{QPPO_}

\begin{algorithm}[t]
\caption{Proposed Q-PPO Algorithm for Solving \eqref{optt}}
{\footnotesize
\begin{algorithmic}[1]
\State \textbf{Input}: Maximum interaction steps, batch size, number of epochs $K$, mini-batch size, discount factor $\gamma$, GAE parameter $\lambda$, clipping threshold $\epsilon$.

\State \textbf{Initialization}: Hybrid quantum–classical actor network $\pi_{\theta_a}$ and classical critic network $V_{\theta_c}$. 
\State Determine the number of sampling iterations $M = \frac{\text{max interaction steps}}{\text{batch size}}$.

\For{$m = 1$ to $M$}
    \State Collect batch size trajectories using policy $\pi_{\boldsymbol{\theta}_{\text{a}}}$;
    \For{each step $t$}
        \State Interact with environment, record $(\boldsymbol{s}_t, \boldsymbol{a}_t, r_t, s_{t+1}, \text{done})$;
    \EndFor
    \State Compute state-value function $V^\pi(\boldsymbol{s}_t)$ using~\eqref{discrew};
    \State Compute GAE $\hat{A}(\boldsymbol{s}_t)$ using~\eqref{advfunc};
    \State Store trajectories, returns, and advantages in buffer
    \For{$k = 1$ to $K$}
        \State Shuffle buffer and divide into mini-batches;
        \For{each mini-batch}
            \State Extract mini-batch data: states, actions, returns, advantages;
            \State Compute policy ratio $p_t({\boldsymbol{\theta}_{\text{a}}}) = \frac{\pi_{\theta_a^{\rm{new}}}(\boldsymbol{a}_t|\boldsymbol{s}_t)}{\pi_{\theta_a^{\rm{old}}}(\boldsymbol{a}_t|\boldsymbol{s}_t)}$;
            \State Compute clipped objective $\mathcal{L}^{\text{CLIP}}(\boldsymbol{\theta}_{\text{a}})$ based on~\eqref{lossactor};
            \State Compute value loss $\mathcal{L}^{\text{VF}}(\boldsymbol{\theta}_{\text{c}})$ using~\eqref{losscritic};
            \State Compute entropy bonus \resizebox{0.45\hsize}{!}{$K(\pi_{\boldsymbol{\theta}_{\text{a}}}) = \mathbb{E}_t \left[ - \sum_a \pi_{\boldsymbol{\theta}_{\text{a}}}(a|s)\log\pi_{\boldsymbol{\theta}_{\text{a}}}(a|s) \right]$};
            \State Compute total loss: $\mathcal{L}$ based on~\eqref{overactorloss};
            \State Update parameters $\boldsymbol{\theta}_{\text{a}}$ and $\boldsymbol{\theta}_{\text{c}}$;
        \EndFor
    \EndFor
\EndFor

\end{algorithmic}
}
\label{Q-PPO}
\end{algorithm}

Building upon the classical PPO framework, the proposed Q-PPO algorithm preserves the standard training pipeline while enhancing the policy representation through the integration of a PQC into the actor network. Specifically, both classical PPO and Q-PPO follow the same overall procedural framework, as summarized in Algorithm~\ref{Q-PPO} and visually illustrated in Fig.~\ref{fig:qppo}. At the beginning of training, a vectorized simulation environment is constructed and reset to generate initial states. The actor-critic networks, together with their corresponding parameters and optimizers, are then initialized. Replay buffers are allocated to store observed states, actions, and rewards. During each training iteration, the agent interacts with the environment to collect transition data, after which returns and advantages are computed based on the recorded rewards, estimated state values, and termination signals. The actor and critic parameters are subsequently updated by minimizing the objective functions in~\eqref{lossactor} and~\eqref{losscritic}, respectively, through an optimizer.

Although both algorithms share the same training procedure and optimization routine, they differ fundamentally in their policy representations. Classical PPO parameterizes the actor using a deep NN, whereas Q-PPO replaces the deep NN with a PQC, thereby enabling quantum-enhanced policy learning. As shown in Fig.~\ref{fig:qppo}, this modification is confined to the actor block, while the critic network remains fully classical, following the framework introduced in~\cite{kwak2021introduction}. Consequently, the principal distinction between classical PPO and Q-PPO lies in the actor's policy model. Furthermore, following the hybrid design philosophy in~\cite{QPPO}, and to address the extremely high dimensionality of SIM configuration as well as the limited scalability of current quantum hardware, we adopt a hybrid quantum–classical architecture that integrates pre-encoding NNs, PQCs, and post-processing NNs. In this design, PQC is responsible for learning compact and expressive policy representations, while lightweight classical modules handle large-scale state and action mappings. This hybrid structure effectively balances representational power, learning efficiency, and hardware feasibility. Therefore, the proposed hybrid quantum-classical Q-PPO framework generally maintains the standard training procedure while introducing a strongly quantum-enhanced policy representation even with limited quantum hardware. Before presenting the detailed PQC architecture, we next provide an overview of how quantum mechanisms enhance the learning process of DRL agents.

\subsubsection{Quantum Computing-enhanced Reinforcement Learning}
In quantum computing, the basic units of information are represented by the computational basis states of a single quantum bit (qubit) $|0\rangle = [1,0]^{\top}$ and $|1\rangle= [0,1]^{\top}$, corresponding to the $0$ and $1$ states of a classical bit.
However, a crucial distinction is that a qubit can exist in a state of quantum superposition. The system comprising $q$ qubits spans a $2^q$-dimensional Hilbert space $\mathcal{H} = (\mathbb{C}^2)^{\otimes q}$, allowing a general quantum state to be represented as a linear combination of the computational basis states~\cite{cherukuri2025q}:
\begin{equation}
    |\psi\rangle = \sum\nolimits_{i=0}^{2^q-1} c_i |\alpha_i\rangle
\end{equation}
where $c_i$ is a complex coefficient representing the occurrence of $|\alpha_i\rangle$, and according to the probability conservation property of quantum dynamics $\sum_{i=0}^{2^q-1}|c_i|^2 = 1$. When applied in computation, quantum states represent inputs in a simultaneous case and thus allow quantum circuits to evaluate multiple inputs. In the context of DRL, this capability allows quantum-enhanced agents to process multiple state-action pairs~\cite{cherukuri2025q}. Grover's search algorithm~\cite{qu2022secure} further exemplifies this quantum advantage, leveraging superposition and amplitude amplification to locate a target state within an unstructured search space of size $Q$ using only $\mathcal{O}(\sqrt{Q})$ queries, a quadratic speedup over the $\mathcal{O}(Q)$ complexity of classical search. This speedup suggests that quantum-enhanced agents can explore large state-action spaces more efficiently and potentially converge to optimal policies faster through appropriate qubit transformations.

These qubit transformations are implemented through quantum operators $\boldsymbol{U}$, referred to as quantum gates, acting on the Hilbert space of the quantum system. Mathematically, the state evolution under a quantum gate is expressed as
\begin{equation}
    |\psi'\rangle = \boldsymbol{U} |\psi\rangle = \sum\nolimits_{i=0}^{2^q-1} c_i \boldsymbol{U}|\alpha_i\rangle.
\end{equation}
In a closed quantum system, all valid transformations are reversible and can therefore be represented by unitary matrices satisfying $\boldsymbol{U}\boldsymbol{U}^\dagger = \boldsymbol{U}^\dagger \boldsymbol{U} = \boldsymbol{I}$. A PQC consists of a sequence of such unitary operations with trainable parameters, forming a quantum model that can be optimized during learning. For a PQC with $q$ qubits, the overall quantum state lies in a Hilbert space of dimension $2^q$. In amplitude encoding, a classical feature vector of dimension $Q$ can be represented using $q=\lceil \log_2 Q \rceil$ qubits, such that $2^q \geq Q$. This property enables PQCs to represent high-dimensional feature spaces using a relatively small number of qubits, which has motivated their use in quantum machine learning and reinforcement learning applications~\cite{paul2024quantum}. As an illustrative example of quantum computational advantages, Shor’s algorithm~\cite{shor1994algorithms} demonstrates that certain computational problems can be solved more efficiently on quantum computers than with known classical approaches. Specifically, Shor’s algorithm exploits quantum superposition and the quantum Fourier transform to factor a $Q$-bit integer in polynomial time with respect to $Q$, in contrast to the sub-exponential but superpolynomial complexity of the classical factoring algorithms. Inspired by these developments, hybrid quantum-classical learning frameworks have recently attracted attention for handling high-dimensional optimization problems. In this work, the PQC is incorporated into the DRL framework to provide a compact and expressive policy representation for the SIM-assisted secure communication problem $(\mathcal{P}_1)$. We also note that the agent in the proposed quantum DRL framework can be trained on classical computers, enhancing practicality in the context of limited quantum hardware resources. Next, we present the PQC architecture in detail.

\subsubsection{Parameterized Quantum Circuit}
In this work, we adopt a widely used hardware-efficient ansatz-based PQC~\cite{kandala2017hardware}. This architecture primarily employs single-qubit Pauli rotation gates and two-qubit entangling gates to construct expressive yet shallow quantum circuits suitable for near-term quantum devices. Specifically, we utilize the Pauli-$Y$ and Pauli-$Z$ operators, whose matrix representations are given by:
\begin{equation}
\boldsymbol{Y}  = 
    \begin{bmatrix}
    0 & -i \\
    i & 0
    \end{bmatrix};
    \quad
    \boldsymbol{Z} =
    \begin{bmatrix}
    1 & 0 \\
    0 & -1
    \end{bmatrix}
\end{equation}
where $i=\sqrt{-1}$. These operators enable rotations of a qubit around the corresponding Bloch sphere axes, thereby controlling both the amplitude and phase of the quantum state. Let $\varphi$ denote the rotation angle. The associated parameterized single-qubit rotation gates are defined as
\begin{equation}
\boldsymbol{R}_Y(\varphi) = \exp\left(-i \varphi \boldsymbol{Y}/2 \right); ~\boldsymbol{R}_Z(\varphi) = \exp\left(-i \varphi \boldsymbol{Z}/2\right).
\end{equation}

To introduce quantum entanglement between qubits, we further employ the two-qubit controlled-$Z$ ($\boldsymbol{\boldsymbol{CZ}}$) gate, which is a standard entangling operator in hardware-efficient PQC designs and is represented by
\begin{equation}
\boldsymbol{\boldsymbol{CZ}} = \mathrm{diag}(1,1,1,-1).
\label{eq:pauli_matrices}
\end{equation}
These parameterized rotation and entangling gates together form the fundamental building blocks of the proposed PQC architecture.

\begin{figure*}[t]
\centering\includegraphics[width=0.8\textwidth]{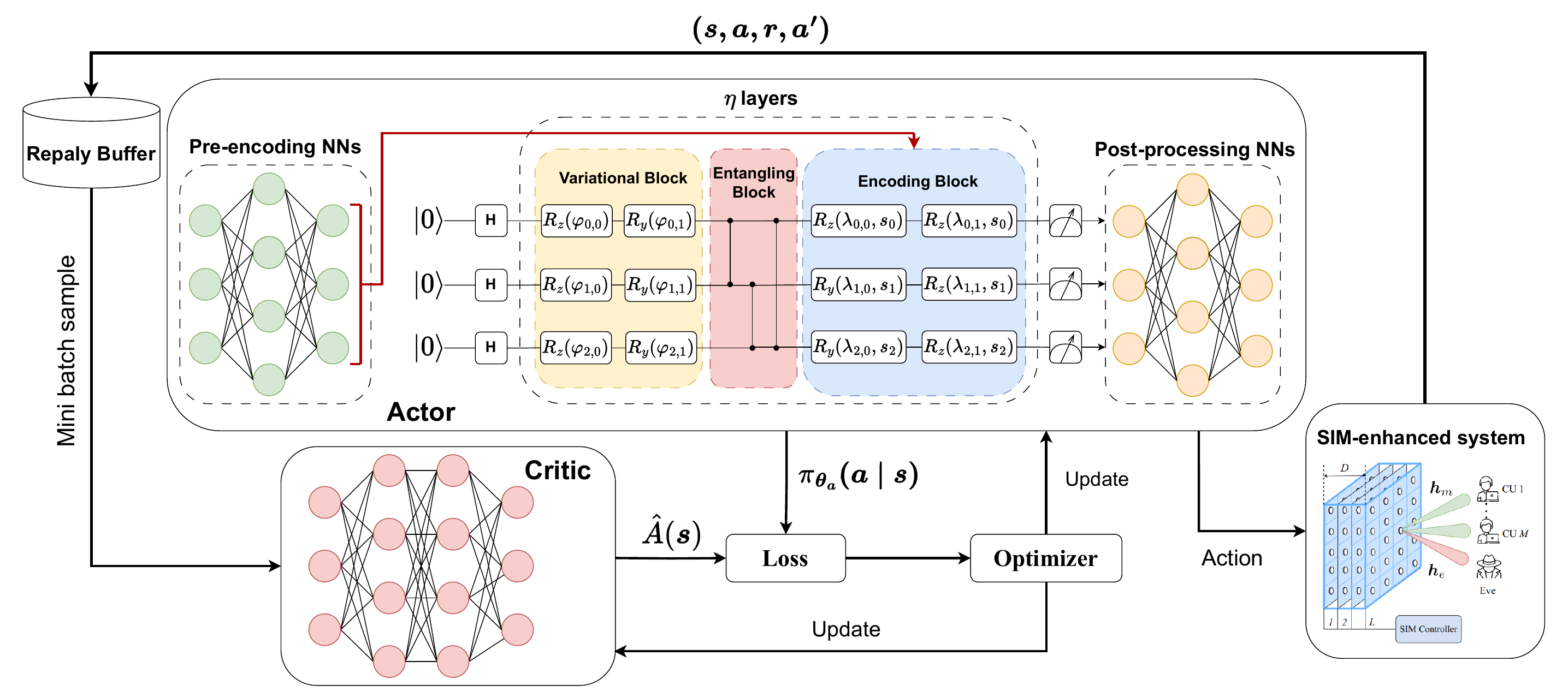}
    \caption{The proposed hybrid quantum-classical Q-PPO framework.}
    \label{fig:qppo}
\end{figure*}

As illustrated in Fig.~\ref{fig:qppo}, the proposed PQC architecture is composed of $\eta$ layers, each containing three blocks: an encoding block, a variational block, and an entangling block. The PQC starts from the fixed initial state ${|0\rangle}^{\otimes q}$, where $q$ denotes the number of qubits. A Hadamard gate ($\boldsymbol{H}$) is first applied to create a superposition state. Next, the vectorized classical medium states $\boldsymbol{s}_t$ are encoded into quantum states $|\boldsymbol{s}_t\rangle$ suitable for processing within the PQC. This encoding is implemented in the encoding block using $\boldsymbol{R}_Y(\upsilon_{i,j,y},s_i)$ and $\boldsymbol{R}_Z(\upsilon_{i,j,z},s_i)$ rotation gates, where $\upsilon_{i,j}$ is the input scaling parameter corresponding to qubit $i$ and layer $j$, and $s_i$ denotes the state feature of the environment encoded by qubit $i$. Since these gates have a periodicity of $2\pi$ while the observable space of the classical medium may exceed this range, the classical states are pre-normalized to the interval $[-\pi, \pi]$ by using the input scaling parameters.
The encoding unitaries in each PQC layer are then given by
\begin{equation}
    \boldsymbol{U}_{\text{enc}}=\bigotimes_{i=0}^{q-1} \boldsymbol{R}_Z(\upsilon_{i,j,z},s_j) \boldsymbol{R}_Y(\upsilon_{i,j,y},s_j).
\end{equation}

The variational and entangling blocks then act similarly to hidden layers in a NN, performing cumulative transformations on the quantum states. Specifically, each $i$-th qubit is rotated by an angle to produce the required values, such as actions or states. In this design, the rotations are performed using $\boldsymbol{R}_Y(\varphi_{i,j,y})$ and $\boldsymbol{R}_Z(\varphi_{i,j,z})$ gates, where $\varphi_{i,j}$ represents the rotation angle of qubit $i$ in layer $j$. The variational unitaries in each PQC layer are then given by
\begin{equation}
    \boldsymbol{U}_{\text{var}}=\bigotimes_{i=0}^{q-1} \boldsymbol{R}_Y(\varphi_{i,j,y}) \boldsymbol{R}_Z(\varphi_{i,j,z}).
\end{equation}
Each layer then applies a sequence of $\boldsymbol{CZ}$ gates to establish entanglement between neighboring qubits, denoted by $\boldsymbol{U}_{\text{ent}}$. This process is repeated across $\eta$ layers using the data re-uploading technique, where the input data are repeatedly encoded throughout the circuit to enhance the expressive capability of the PQC. The actor PQC unitary map with $\eta$ layers could be written compactly as
\begin{equation}
    \boldsymbol{U}(\boldsymbol{\theta}_a) = \prod_{j=1}^{\eta} U_{\text{ent}} \boldsymbol{U}_{\text{var}}(\boldsymbol{\varphi}) \boldsymbol{U}_{\text{enc}}(\boldsymbol{\upsilon, \boldsymbol{s}})
\end{equation}
where $\boldsymbol{\theta}_a \triangleq (\boldsymbol{\upsilon},\boldsymbol{\varphi})$ are trainable parameters.

\subsubsection{Quantum Measurement}
The output of the PQC is obtained through quantum measurements performed on the qubits. A projective measurement is described by a Hermitian observable $O$ acting on the state space of the quantum system, where the observable defines the eigenbasis associated with the measurement. The measurement outcomes correspond to the eigenvalues $\alpha_m$ obtained from the spectral decomposition of $O$, while the corresponding post-measurement states are determined through the orthogonal projection operators $P_m$. According to the Born rule, when a quantum state $|\psi\rangle$ is measured, it produces an outcome $\alpha_m$ and is projected onto the state 
$P_m |\psi\rangle/\sqrt{p(m)}$. 
The probability of obtaining this outcome is $p(m) = \langle \psi | P_m | \psi \rangle = \langle P_m \rangle_{\psi}$. It can be seen that for actor networks, this measurement process is equivalent to performing an action $\boldsymbol{a}$ under policy $\pi_{\boldsymbol{\theta_a}}(\boldsymbol{a}\mid \boldsymbol{s})$.
Consequently, the quantum policy can be defined as
\begin{equation}
\pi_{\boldsymbol{\theta_a}}(\boldsymbol
{a}\mid \boldsymbol{s}) = \langle P_{\boldsymbol{a}} \rangle_{\boldsymbol{s}, \boldsymbol{\theta_a}}.
\end{equation}

However, the policy defined directly from projective measurements can only address a discrete set of actions, whereas the power allocation and SIM phase shifts form a continuous action space. We therefore retain the PQC as a state-dependent feature map and place the stochastic policy on its measured expectation values rather than on the measurement outcomes themselves. A fixed set of $q$ Hermitian observables $\{\hat{O}_i\}_{i=1}^{q}$ is
measured by introducing trainable output-scaling weights $\boldsymbol{w}$~\cite{QPPO}:
\begin{equation}
z_i(\boldsymbol{s};\boldsymbol{\theta}_a)
=
w_i
\left\langle
0^{\otimes q}
\left|
U(\boldsymbol{s},\boldsymbol{\upsilon},\boldsymbol{\varphi})^{\dagger}
\hat{O}_{i}\,
U(\boldsymbol{s},\boldsymbol{\upsilon},\boldsymbol{\varphi})
\right|
0^{\otimes q}
\right\rangle
\end{equation}
for $i=\{1,\dots,q\}$, where the full set of trainable parameters is
$\boldsymbol{\theta}_a=(\boldsymbol{\upsilon},\boldsymbol{\varphi},\boldsymbol{w})$. The resulting vector $\boldsymbol{z}=[z_1,\dots,z_q]^{\top}$ is passed to the Post-NN, which outputs the mean and the standard deviation of the Gaussian policy from which the transmit power allocation and SIM phase shift actions are sampled.

\subsubsection{Pre-Encoding and Post-Processing}
In the proposed PQC-based actor network, environmental state features are embedded into qubits through parameterized single-qubit rotation gates, specifically $\boldsymbol{R}_Y(\upsilon_{i,j,y}, s_i)$ and $\boldsymbol{R}_Z(\upsilon_{i,j,z}, s_i)$, where the classical state variables are encoded into the rotation angles of the quantum gates following the angle encoding scheme. For continuous multi-dimensional control problems, each state feature $s_i$ corresponds to one dimension of the state vector $\mathbf{s}$ and can be mapped to a single qubit. Consequently, the number of required qubits must be at least equal to the dimensionality of the state–action space. 

In low-dimensional quantum DRL tasks, such direct encoding into rotation gates is effective~\cite{QPPO}. However, for large-scale SIM optimization, the state and action spaces are continuous and high-dimensional, making direct encoding challenging under limited qubit resources. Moreover, increasing circuit depth improves learning performance only up to a certain threshold~\cite{skolik2022quantum}, since excessively deep PQCs may suffer from barren plateaus, where gradients vanish and the training process becomes ineffective~\cite{stkechly2024}. As a result, further increasing the PQC depth may lead to optimization stagnation with limited performance improvement.
These constraints motivate the integration of classical NNs to assist PQCs in processing complex state and action spaces.

Classical NN architectures are constructed by cascading multiple transformation layers, which can be represented as a composition 
\begin{equation}
    \mathcal{Q} = \mathcal{L}_{b_{\eta-1} \to b_{\eta}} \circ \mathcal{L}_{b_{\eta-2} \to b_{\eta-1}} \circ \cdots \circ \mathcal{L}_{b_0 \to b_1}
\end{equation}
where $\circ$ denotes the function composition operator, and $\eta$ denotes the number of layers, and the subscripts indicate the input and output dimensions of each layer. Each layer maps a $b$-dimensional input to a $b'$-dimensional output through operations such as linear transformations followed by nonlinear activations, which can be expressed as
\begin{equation}
\mathcal{L}_{n_0 \to n_1}(\mathbf{x}) = \sigma(\boldsymbol{\theta}\mathbf{x} + \boldsymbol{\rho})
\end{equation}
where $\mathbf{x}$ is an input vector of dimension $b$, $\boldsymbol{\theta}$ is a matrix of size $b' \times b$, and $\boldsymbol{\rho}$ is a vector of size $b'$, the output of a single layer $\mathcal{L}_{b \to b'}$. 

Leveraging this hierarchical representation capability, the proposed hybrid quantum architecture integrates classical NNs both before and after the PQC~\cite{QPPO}. Specifically, the Pre-NN adopts the efficient dimensionality-reduction strategy in~\cite{mnih2013playing}, where convolutional layers pre-encode high-dimensional observed states into compact feature representations, which are subsequently mapped to PQC encoding gates through a fully connected layer. This provides a rich yet compact representation compatible with the limited number of available qubits. Similarly, the Post-NN comprises fully connected layers applied to quantum measurement outputs to generate continuous multi-dimensional actions. The Post-NN compensates for the limited expressiveness of shallow PQCs constrained by the number of qubits and enhances the overall representational capacity of the hybrid policy network. Unlike conventional DRL architectures that rely on deep and wide NNs, the proposed Pre-NN/Post-NN modules employ only a small number of hidden layers and neurons, thereby avoiding excessive classical computational overhead, as further discussed in Section~\ref{Sec:simulation}.

\section{Performance Evaluation}
\label{Sec:simulation}
In this section, we evaluate the performance of the proposed Q-PPO algorithm and compare it with several benchmark schemes, including PPO, DDPG, TD3, and a random action selection strategy. The simulation setup and parameter configurations are detailed below.

\subsection{Parameter Settings}
We consider a SIM-assisted downlink system, whose system configuration is adopted from~\cite{10922857,kavianinia2025secrecy} and~\cite{10949617}. Specifically, the SIM is integrated into the BS and deployed at a height of $H_{\mathrm{b}} = 10$ m. The distance from the antenna array to the output metasurface is set to $D = 5\lambda$. The BS transmit power is $P_0 = 10$ dBm, while the noise powers at the legitimate users ($\sigma_m^2$) and the eavesdropper ($\sigma_e^2$) are both set to $-104$ dBm. {The number of CUs is set to $M=3$.} {CUs are randomly distributed within an annular region centered at the BS, with inner and outer horizontal-distance limits $D_{\min}$ and $D_{\max}$, respectively. Under the adopted deployment geometry, the propagation distance from the output metasurface to CU $m$ or Eve is $d_{\varepsilon}=\sqrt{H_{\mathrm{b}}^2+D_{\varepsilon}^2}$, where $\varepsilon=m$ denotes CU $m$, $\varepsilon=e$ denotes Eve, and $D_{\varepsilon}$ is the horizontal distance to the corresponding node. The distance parameters $D_{\min}$ and $D_{\max}$ are measured in metres, whereas the minimum QoS rate $R_{\min}$ is measured in bits/s/Hz.} For the channel model, the path loss at a reference distance of 1 m is $C_0 = -35$ dB, and the path loss exponent is $\alpha = 3.5$. The variance of the CSI estimation error is given by $\delta_{e}^2 = \frac{\delta^2}{N}\|\hat{\boldsymbol{h}}_{e}\|_2^2$~\cite{9374975}, where $\delta = 0.1$ represents the relative level of CSI uncertainty. 

The main hyperparameters of the quantum and classical PPO algorithms follow the standard configuration of the PPO implementation in~\cite{stable-baselines3}, including a learning rate ($lr$) of $3 \times 10^{-4}$, a clipping parameter of $0.2$, a batch size of 64, a discount factor of $0.99$, and a GAE parameter of $0.95$ for the bias–variance trade-off. The classical actor–critic network consists of four hidden layers with 1024 neurons each, employing ReLU activation and the Adam optimizer. In the hybrid classical–quantum architecture, the Pre-NN contains two convolutional layers with 128 neurons and one fully connected layer with $64$ neurons, while the Post-NN includes two fully connected layers with $62$ and $32$ neurons, respectively. The PQC comprises four layers operating on five qubits. As described in Section \ref{QPPO_}, both the classical and quantum PPO share the same training process, where weight updates are performed every $1024$ steps using the Adam optimizer. To enable sufficient exploration, the environment is updated every $20$ steps. The remaining simulation parameters are summarized in Table~\ref{Table:parameter}.

\begin{table}[t]
    \centering
    \caption{Simulation Parameters}
    \begin{tabular}{|c|c||c|c|}
    \hline
    Parameter & Value & Parameter & Value\\
    \hline
    $L$&  $3$ & $N$ & $25$ \\
    $d_{\mathrm{a}}$ & $\lambda/2$~\cite{kavianinia2025secrecy}  & $\lambda$ & $10.7 $mm~\cite{10949617}  \\
    $A_t$ & $\lambda^2/4$& $\kappa$ & $-30$ dB~\cite{10949617} \\
    {$D_{\min}$} & $75$ m& {$D_{\max}$} & $100$ m \\
    $c_1$ & $0.01$~\cite{stable-baselines3} & $c_2$ & $0.5$~\cite{stable-baselines3} \\
    \hline
    \end{tabular}
    \label{Table:parameter}
\end{table}

\subsection{Complexity Analysis}
In this subsection, we provide a comparative complexity analysis between the proposed Q-PPO framework and the standard PPO baseline, focusing on the number of trainable parameters and the computational complexity of the actor networks. Since the PQC is integrated only into the actor component, the analysis focuses on comparing the hybrid quantum-classical actor with its fully classical counterpart.

\textit{1) Number of trainable parameters:}
    In a standard fully connected neural network, the trainable parameters consist of weights and biases. For two adjacent layers, the number of trainable parameters is given by~\cite{van2024dynamic} $(|\mathcal{A}_i|+1)|\mathcal{A}_{i+1}|$, where $|\mathcal{A}_i|$ denotes the number of neurons in the $i$-th layer. Accordingly, the total number of trainable parameters in the classical actor network is expressed as
   \begin{equation}
        P_{\text{NN}} = \sum\nolimits_{i=1}^{\eta_{\text{NN}}-1} (|\mathcal{A}_i| + 1) \times |\mathcal{A}_{i+1}|
    \end{equation}
    where $\eta_{\text{NN}}$ denotes the total number of layers.

    For the PQC, the trainable parameters are defined as $\boldsymbol{\theta}_a=(\boldsymbol{\upsilon},\boldsymbol{\varphi},\boldsymbol{w})$. Specifically, the encoding parameters $\boldsymbol{\upsilon}$ and the variational rotation parameters $\boldsymbol{\varphi}$ each contribute $2\eta q$ parameters, while the measurement weights $\boldsymbol{w}$ contribute $q$ parameters. Thus, the total number of trainable parameters in the PQC is $P_{\text{PQC}} = (4\eta + 1)q.$ Consequently, the total number of trainable parameters in the hybrid quantum-classical actor is given by
        \begin{equation}
            P_{\text{quantum}} = P^{\text{Pre}}_{\text{NN}} + P^{\text{Post}}_{\text{NN}} + P_{\text{PQC}}.
        \end{equation}

\textit{2) Computational complexity:}
    The computational complexity of a forward pass through the classical neural network is proportional to the total number of multiplications between consecutive layers. Therefore, the overall complexity of the classical PPO actor is expressed as
    \begin{equation}
        \mathcal{O}\Big(T N_b \big(\sum\nolimits_{i=1}^{\eta_{\text{NN}}-1} |\mathcal{A}_{i}| \times |\mathcal{A}_{i+1}| \big)\Big)
    \end{equation}
  where $N_b$ denotes the training batch size and $T$ is the total number of training iterations. 

In the PQC, each circuit layer contains $4q$ parameterized single-qubit rotation gates associated with the encoding and variational blocks. Assuming each gate operation contributes one elementary operation, the complexity per circuit layer is $\mathcal{O}(4q)$. In addition, Pauli-$Z$ expectation measurements are performed on all $q$ qubits, resulting in measurement complexity $\mathcal{O}(q)$. During training, the parameter-shift rule is employed for gradient evaluation~\cite{11389788}, requiring two forward evaluations for each trainable gate parameter. Accordingly, the total complexity associated with PQC parameter updates is given by $\mathcal{O}\left(2(4\eta + 1)q\right).$ Then, the total complexity of the proposed Q-PPO can be expressed as
\begin{align}
\mathcal{O}\Big(
TN_b \big(&2(4\eta+1)q+\sum\nolimits_{i=1}^{\eta_{\text{Pre}}-1} |\mathcal{A}^{\text{Pre}}_{i}| |\mathcal{A}^{\text{Pre}}_{i+1}| \nonumber\\
&+ \sum\nolimits_{j=1}^{\eta_{\text{Post}}-1} |\mathcal{A}^{\text{Post}}_{j}| |\mathcal{A}^{\text{Post}}_{j+1}|\big)\Big).
\end{align}

In our implementation, the input dimension is $2N(M+1)=200$, corresponding to the CSI of the CUs and the imperfect CSI of Eve, while the output dimension is $LN+M=78$, corresponding to power allocation and SIM phase shift configurations. The resulting parameter counts and computational complexity comparisons are summarized in Table~\ref{tab:complexity_comparison}. The results show that the proposed hybrid Q-PPO framework achieves improved secrecy performance and convergence behavior while using substantially fewer trainable parameters than the classical PPO baseline.
\begin{table}[t]
\centering
\caption{Complexity Comparison between Classical and Hybrid Quantum Actor}
\label{tab:complexity_comparison}
\setlength{\tabcolsep}{8pt}
\begin{tabular}{lrr}
\toprule
\textbf{Component} & \textbf{Parameters} & \textbf{FLOPs} \\
\midrule
Classical Actor & 3{,}434{,}574 & $TN_b(3.43 \times 10^6)$ \\
\midrule
\quad Pre-NN & 50{,}821 &  \\
\quad PQC & 85 &  \\
\quad Post-NN & 7{,}374 &  \\
\textbf{Hybrid Actor} & \textbf{58{,}280} & $\boldsymbol{TN_b(4.13 \times 10^4)}$ \\
\bottomrule
\end{tabular}
\end{table}

\subsection{Baselines and Performance Metrics}
\noindent \textit{\textbf{Baselines}:}
To demonstrate the adaptability of the proposed Q-PPO algorithm and its effectiveness in enhancing SIM-enabled secure communications, we compare its performance with the following most relevant baseline schemes:
\begin{itemize}
\item \textbf{PPO:} The conventional PPO algorithm~\cite{ppo} is implemented to solve the optimization problem in \eqref{optt}. This baseline serves to isolate and quantify the performance gains achieved by incorporating quantum-enhanced policy representation.
\item \textbf{TD3 and DDPG:} These baselines are implemented using state-of-the-art DRL algorithms that are widely adopted for continuous control problems~\cite{10949617,yang2024}, providing a fair comparison with established learning-based SIM optimization approaches.
\item \textbf{Random:} A random configuration scheme is also considered, where the SIM phase shifts are randomly selected within their feasible ranges. This baseline highlights the importance of intelligent SIM control in achieving secure communication performance.
\end{itemize}

\vspace{0.1cm}
\noindent \textit{\textbf{Performance metrics}:}
We evaluate the proposed quantum-DRL-based joint resource allocation and SIM design scheme using two key performance metrics, including the ASR and Jain’s fairness index (JFI)~\cite{jain}. {The ASR, measured in bits/s/Hz, serves as the objective of problem~$(\mathcal{P}_1)$ and reflects the effectiveness of the joint power allocation and SIM design. The provided ASR values are averaged over all sampled realizations, with QoS-violating realizations contributing zero in accordance with~\eqref{reward}. Specifically, Figs.~\ref{impact1} and \ref{fig:impact_params} illustrate the performance during training, while Figs.~\ref{fig:Performance Comparison} and \ref{fig:tradeJFI} present the results after policy convergence.
A higher ASR indicates better secrecy performance. To evaluate fairness among the CUs, we adopt JFI ~\cite{jain}, defined as
\begin{equation}
    J = \frac{\Big( \sum_{m=1}^{M} R_m \Big)^{2}}{M \sum_{m=1}^{M} R_m^{2}}.
\end{equation}
For nonnegative secrecy rates with at least one positive value, $J \in [1/M,\,1]$, where \(1/M\) indicates no fairness and $``1"$ represents perfect fairness.}

\subsection{Results and Discussion}
\subsubsection{Convergence Analysis and Impact of Imperfect CSI}
\begin{figure}[t]
    \centering
    \subfloat[Performance comparison among Q-PPO, PPO, DDPG, TD3, and Random schemes\label{fig:reward}]
    {\includegraphics[width=0.75\linewidth]{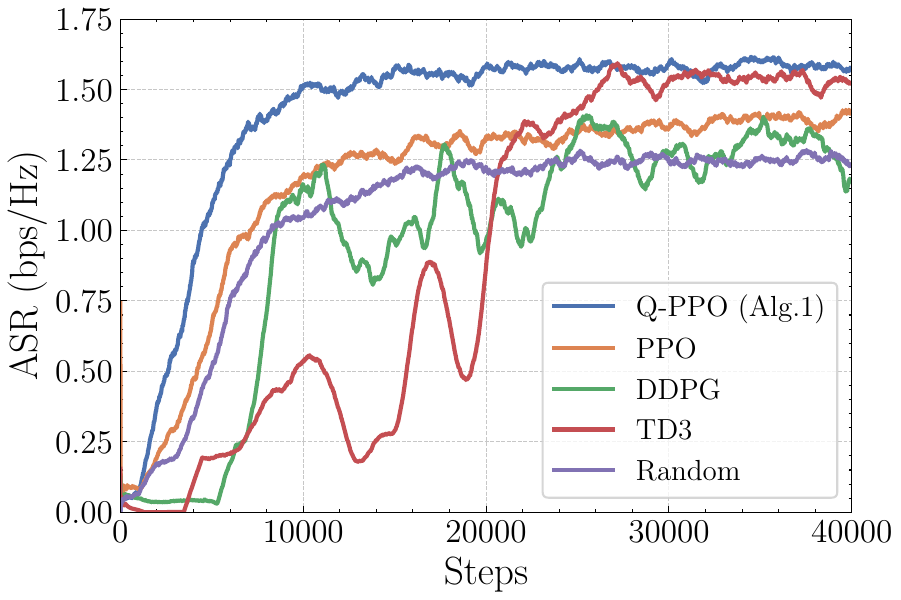}}\hfill
    \subfloat[Influence of the CSI uncertainty level ($\delta$) on ASR evolution\label{fig:imcsi}]
    {\includegraphics[width=0.75\linewidth]{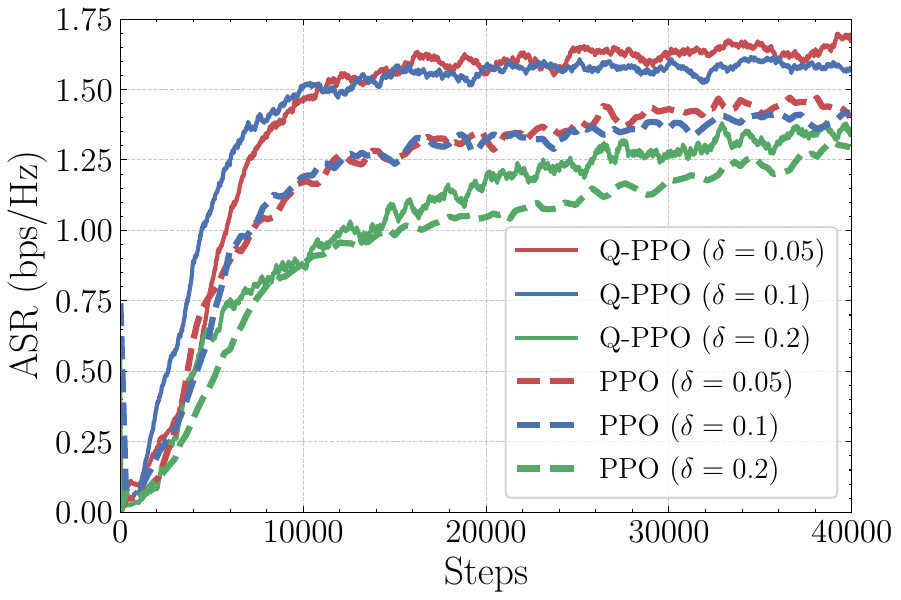}}\hfill
    \caption{(a) Convergence performance comparison of different algorithms, and (b) Impact of the CSI uncertainty level ($\delta$) on the convergence behavior.}
    \label{impact1}
\end{figure}

We first evaluate the convergence behavior of the different algorithms. As shown in Fig.~\ref{fig:reward}, all algorithms converge steadily toward stable reward values, confirming their effectiveness in solving the optimization problem. As expected, the PPO-based schemes (including the conventional PPO and our proposed Q-PPO algorithm) achieve more stable convergence behavior than that of the other methods. Notably, the proposed Q-PPO algorithm exhibits superior performance with significantly faster convergence, achieving an optimal ASR of $1.67$ bps/Hz, approximately $15\%$ higher than that of the conventional PPO, after more than $20,000$ training steps. In contrast, the conventional PPO reaches performance saturation only after nearly $30,000$ training steps, indicating that Q-PPO achieves approximately a $30\%$ improvement in convergence speed. Although the advanced TD3 algorithm performs closely to Q-PPO, it similarly requires nearly $30,000$ steps to converge. This highlights the effectiveness of the PQC in enhancing the agent’s ability to explore the environment through quantum advantages such as improved state representation and exploration efficiency. Furthermore, comparison with the Random scheme reveals that the SIM phase shift design improves system security performance by approximately $22\%$.

\begin{figure*}[t]
    \centering
    \subfloat[ASR convergence under different learning rates, $lr$\label{fig:lr}]
    {\includegraphics[width=0.3\linewidth]{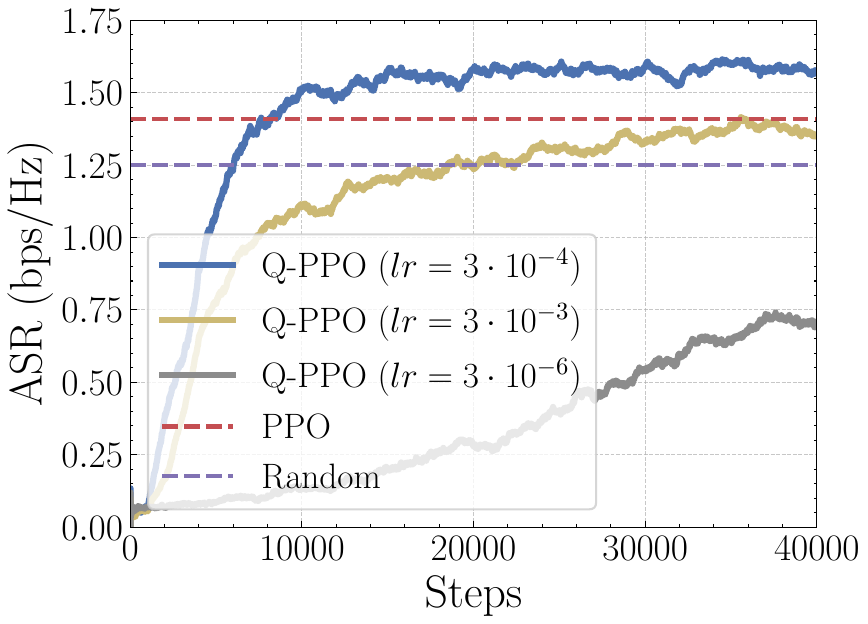}}\vspace{0.1cm}
    \subfloat[ASR convergence under different numbers of qubits, $q$ \label{fig:qubit}]
    {\includegraphics[width=0.3\linewidth]{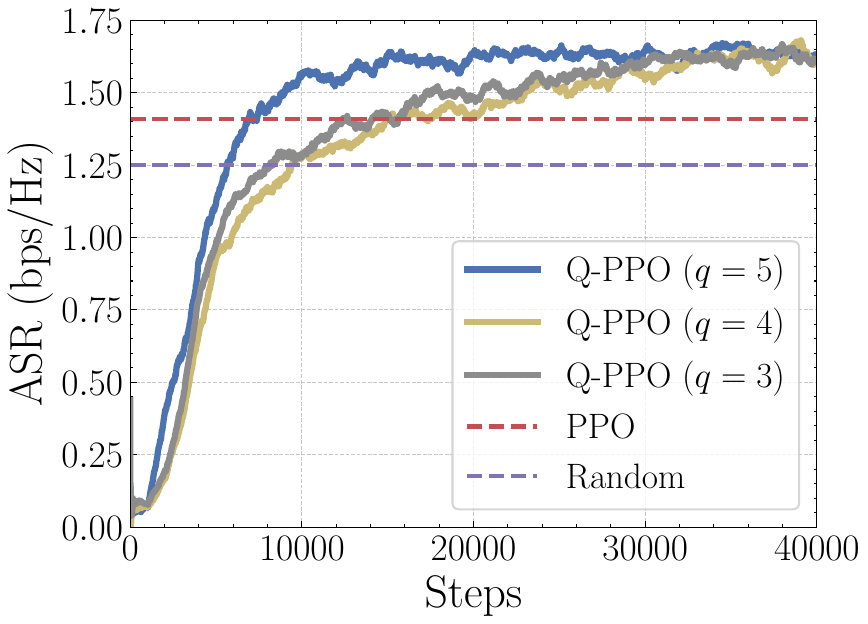}}\vspace{0.1cm}
    \subfloat[ASR convergence under different numbers of PQC layers, $\eta$ \label{fig:qlayer}]
    {\includegraphics[width=0.3\linewidth]{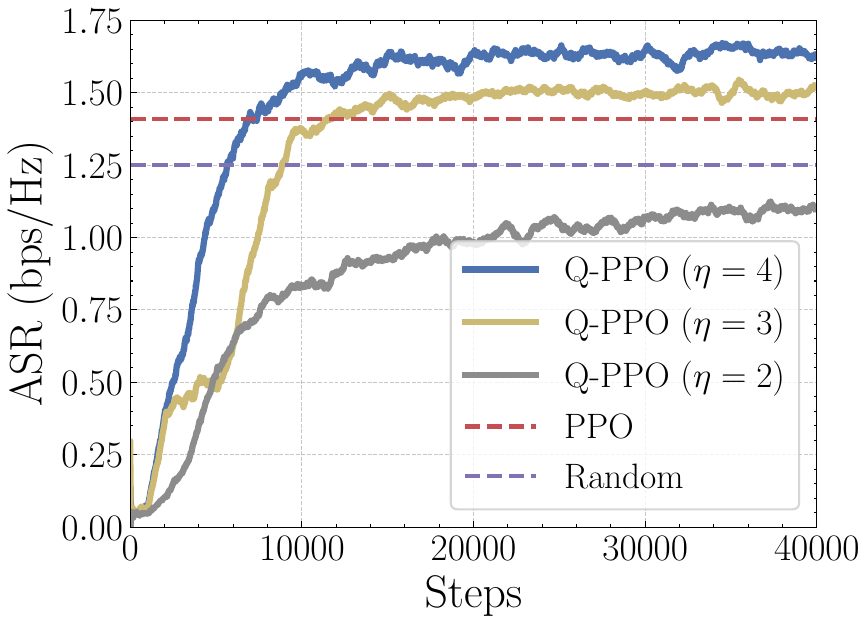}}
    \caption{Convergence performance of the proposed Q-PPO scheme under different hyperparameter settings: (a) learning rate $lr$, (b) number of available qubits $q$ and (c) number of PQC layers $\eta$.}
    \label{fig:impact_params}
\end{figure*}

We next examine the impact of the CSI uncertainty level ($\delta$) on the convergence behavior of the proposed Q-PPO algorithm and the PPO baseline. As illustrated in Fig.~\ref{fig:imcsi}, the ASR consistently improves throughout the training process across all considered configurations, confirming the stable convergence of both frameworks. At lower uncertainty levels ($\delta = 0.05$ and $\delta = 0.1$), both algorithms achieve stable convergence with relatively high ASR performance, indicating that the learned policies can effectively exploit the available CSI information to jointly optimize the SIM phase shifts and transmit power allocation. In contrast, a higher uncertainty level ($\delta = 0.2$) leads to noticeable performance degradation in both frameworks, with the achieved ASR decreasing to approximately $1.35$ bps/Hz for Q-PPO and $1.25$ bps/Hz for PPO. This degradation occurs because imperfect Eve CSI reduces the reliability of the environmental information available to the agent, making it more difficult to accurately characterize the eavesdropping channel conditions. Consequently, the learned policy requires more extensive exploration before converging to a stable strategy, resulting in slower convergence behavior under highly uncertain environments. In addition, under increased CSI uncertainty, the learned policy exhibits more conservative power allocation and more adaptive SIM phase shift configurations in order to maintain secrecy performance despite the unreliable eavesdropper information. Despite the more challenging learning environment, Q-PPO consistently maintains a higher ASR floor than the conventional PPO framework across all uncertainty levels, demonstrating improved robustness in high-dimensional SIM optimization scenarios.

\subsubsection{Impact of Q-PPO Hyperparameters}
We investigate the impact of key hyperparameters of Q-PPO, including learning rate $lr$, the number of available qubits $q$, and the number of PQC layers $\eta$, as illustrated in Fig.~\ref{fig:impact_params}. As shown in Fig.~\ref{fig:lr}, we evaluate different learning rates $lr = ({3\times10^{-3}, 3\times10^{-4}, 3\times10^{-6}})$. The best performance is achieved at $lr = 3\times10^{-4}$, where the algorithm converges faster and achieves higher rewards. This is because a relatively smaller learning rate stabilizes parameter updates and reduces gradient noise, allowing the model to fine-tune its parameters to get a better solution toward a global optimum. However, when the learning rate is too small (\textit{e.g.}, $lr  = 3\times10^{-6}$), Q-PPO is not able to escape from the local optima, resulting in slower learning performance. Conversely, a too large learning rate leads to unstable updates and hinders the model’s ability to fine-tune its policy effectively.

As shown in Fig.~\ref{fig:qubit}, we compare training performance on PQC configurations with $q = \{3,4,5\}$ qubits. It can be seen that the agent converges to the optimal reward after about $20,000$ steps with a PQC of $5$ qubits, while the $3$- and $4$-qubit configurations require about $40,000$ and $35,000$ steps, respectively. Increasing the number of qubits not only expands the encoding block but also increases the number of measurements that can be performed. This allows PQC to encode richer representations of the classical environment and explore a wider action space at each time slot, thereby improving training.

In Fig.~\ref{fig:qlayer}, we vary the number of PQC layers $\eta$, considering $\eta=\{2,3,4\}$, and compare the resulting convergence. We can see that the convergence of the algorithm improves as the number of layers increases. The training process in PQC is performed on each layer, similar to NN, where the trainable parameters are tuned to optimize the objective function. In general, adding more layers to the PQC architecture increases the number of trainable parameters and leads to higher computational complexity. However, this allows the computation on each qubit in PQC to be maximized through data reloading between layers, resulting in improved performance.

\subsubsection{Performance Comparison}
\begin{figure*}[t]
    \centering
    \subfloat[ASR versus  $N$\label{fig:impact meta atom}]
    {\includegraphics[width=0.3\textwidth]{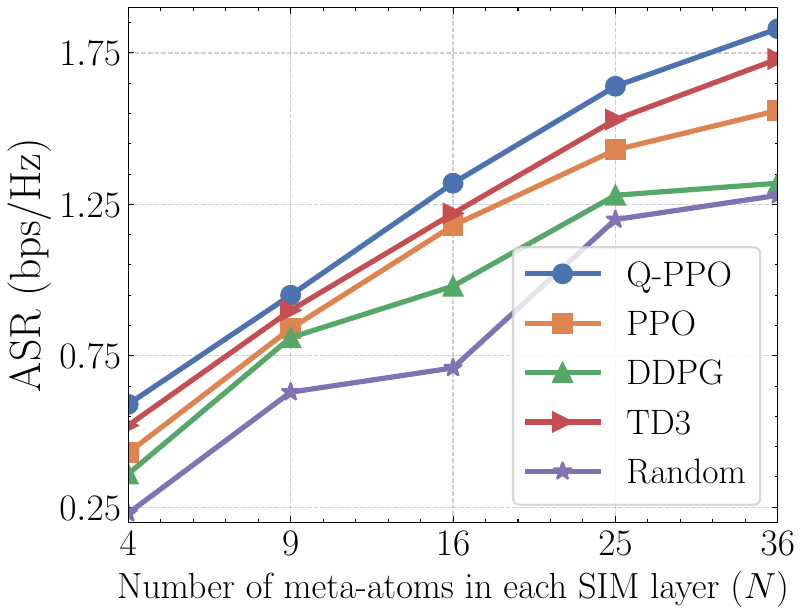}} 
    \hspace{0.3cm}
    \subfloat[ASR versus  $L$\label{fig:impact SIM layers}]
    {\includegraphics[width=0.3\textwidth]{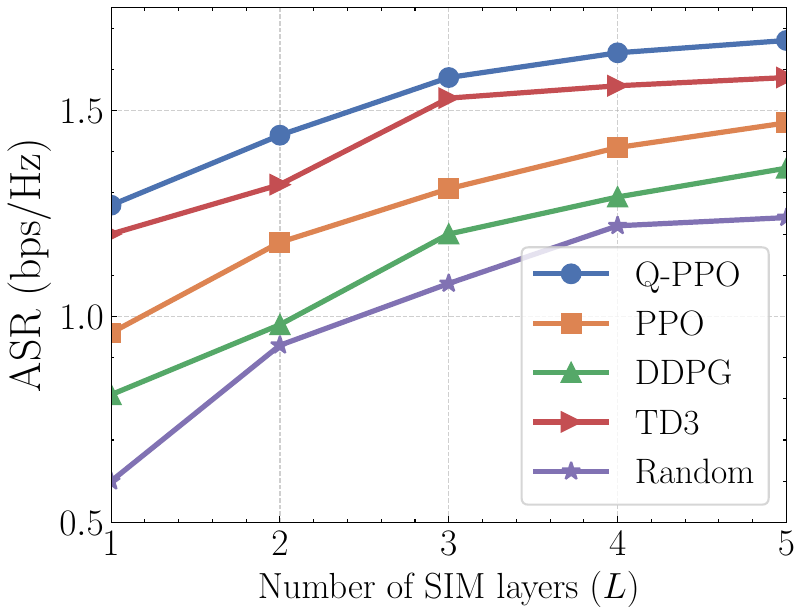}}
    \hspace{0.3cm}
    \subfloat[ASR versus  $P_0$\label{fig:impact transmit power}]
    {\includegraphics[width=0.3\textwidth]{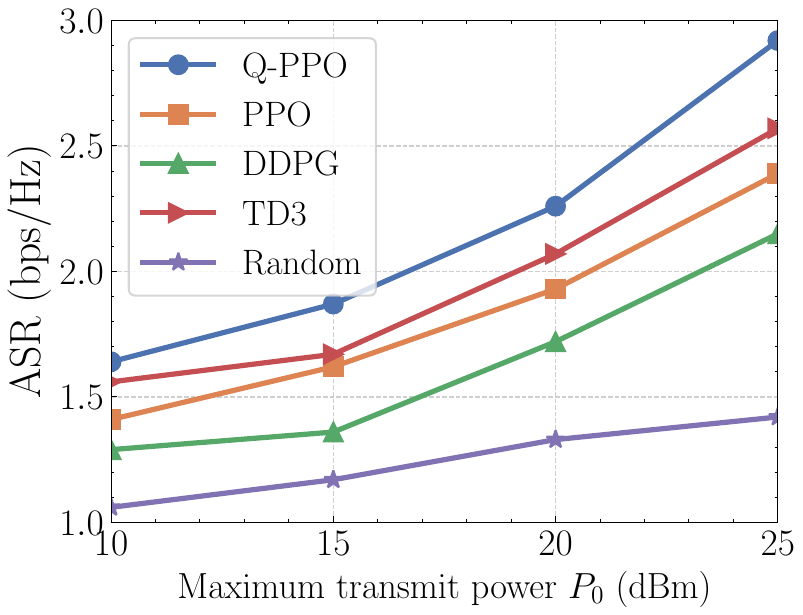}}\\
    \subfloat[ASR versus $R_{\min}$\label{fig:impact data rate requirement}]
    {\includegraphics[width=0.3\textwidth]{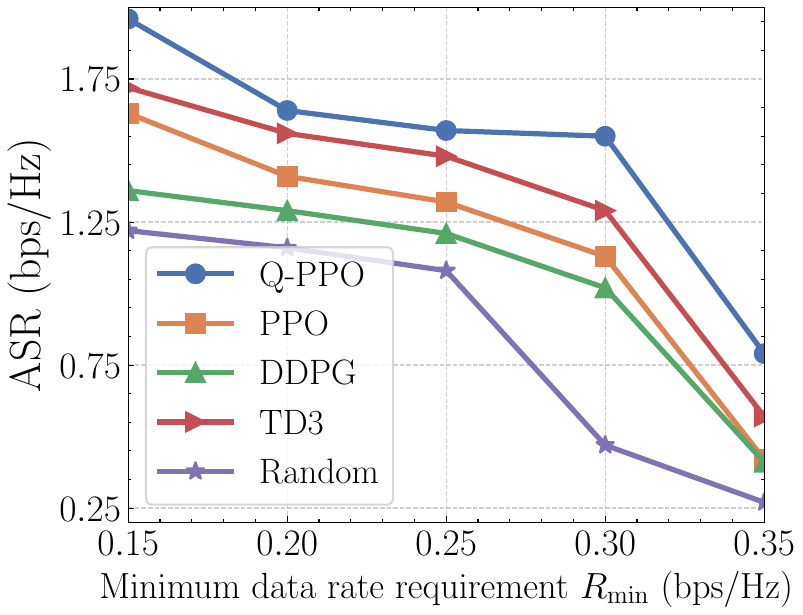}}
    \hspace{0.3cm}
    \subfloat[ASR versus $M$\label{fig:impact number of CUs}]
    {\includegraphics[width=0.3\textwidth]{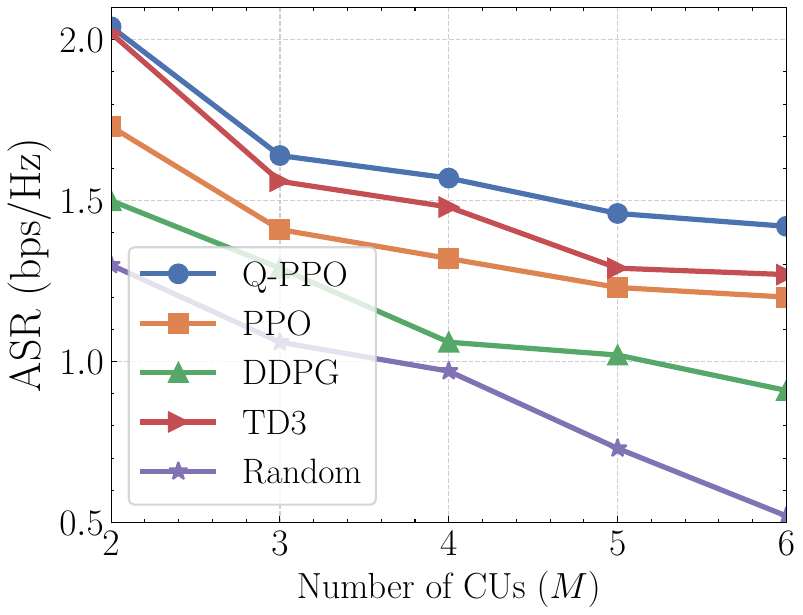}}
    \hspace{0.3cm}
    \subfloat[ASR versus $d_{\varepsilon}$\label{fig:impact distance}]
    {\includegraphics[width=0.3\textwidth]{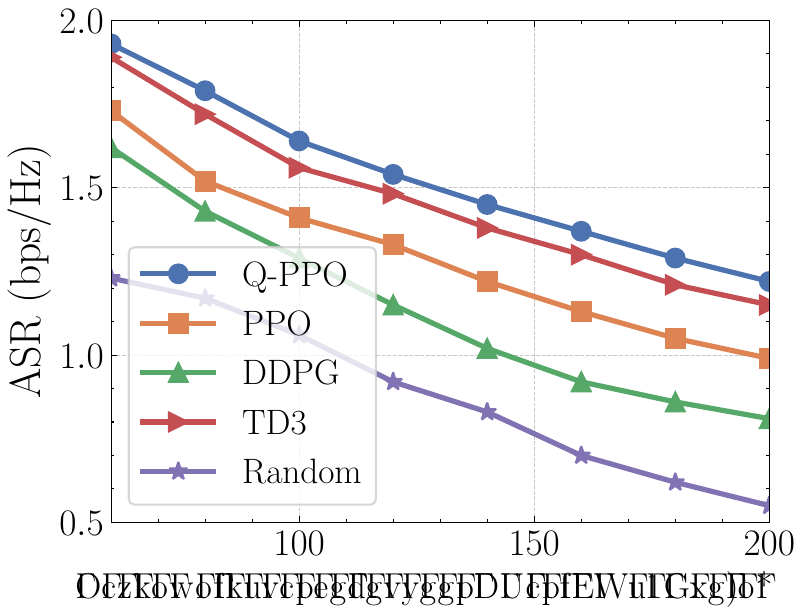}}   
    \caption{Evaluation of the ASR with varying (a) number of meta-atoms per layer ($N$), (b) number of SIM layers ($L$), (c) maximum transmit power ($P_0$), (d) minimum data rate requirement ($R_{\min}$), (e) number of CUs ($M$), and (f) CUs/Eve distance ($d_{\varepsilon}$).}
    \label{fig:Performance Comparison}
\end{figure*}
Fig.~\ref{fig:impact meta atom} illustrates the ASR versus the number of meta-atoms per layer ($N$). As observed, all evaluated schemes exhibit an increasing secrecy rate when $N$ increases. This trend results from the expanded degrees of freedom, enabling the SIM to be more flexibly tuned for better channel conditions. The proposed Q-PPO algorithm consistently achieves the best performance. When each layer contains relatively few meta-atoms (\textit{e.g.}, $N=4$ or $N=9$), PPO and TD3 yield a similar performance to that of Q-PPO due to their lower computational demand. However, as the number of meta-atoms increases, the performance gap becomes more evident, highlighting the PQC-based agent’s enhanced ability to represent the environment and facilitate more efficient policy learning. From a scalability perspective, this trend highlights the robustness of Q-PPO in handling the rapidly expanding action space induced by large-scale SIM architectures. In contrast, conventional DRL methods struggle to maintain performance as the problem dimensionality grows.

Fig.~\ref{fig:impact SIM layers} shows how the ASR varies with the number of SIM layers ($L$). Increasing $L$ consistently improves secrecy performance for all algorithms, confirming the advantage of multilayer SIM architectures in enhancing signal obfuscation and enabling more effective physical-layer precoding. Notably, the proposed Q-PPO algorithm achieves comparable security levels with far fewer SIM configurations. For instance, Q-PPO attains an ASR of $1.3$ bps/Hz using a single SIM layer, whereas DDPG requires four layers to achieve a similar performance. Moreover, despite the growth in the number of meta-atoms that must be designed as $L$ increases, the Q-PPO framework shows superiority in terms of ASR, demonstrating strong scalability. However, as the SIM depth increases, the improvement in secrecy rate gradually saturates due to the rapidly growing number of meta-atoms to be optimized, which substantially increases training complexity.

Fig.~\ref{fig:impact transmit power} illustrates the impact of the transmit power budget $P_0$ on the ASR of the system. As shown, increasing $P_0$ significantly enhances the ASR. Specifically, for the proposed Q-PPO algorithm, the secrecy rate rises from $1.64$ bps/Hz to $2.93$ bps/Hz when the maximum transmit power increases from $10$ to $25$ dBm. This improvement arises because a higher transmit power enables the BS to deliver stronger signals to the CUs.

We examine the effect of the minimum data rate requirement $R_{\min}$ on the ASR, as illustrated in Fig.~\ref{fig:impact data rate requirement}. As $R_{\min}$ increases, tighter QoS constraints lead to a noticeable decline in secrecy performance across all evaluated schemes. In particular, when $R_{\min} = 0.35$ bps/Hz, all schemes experience a sharp degradation, indicating that the system struggles to suppress eavesdropping channels under limited transmit power resources while maintaining the QoS requirements of CUs. Notably, among all compared schemes, the proposed Q-PPO algorithm consistently achieves the highest secrecy rate for various $R_{\min}$ values, demonstrating its strong capability to maintain secure communication even under stringent QoS constraints.

{Fig.~\ref{fig:impact number of CUs} depicts the ASR versus the number of legitimate users ($M$). Since $K=M$, increasing $M$ also increases the number of BS antennas. The observed ASR reduction reflects two factors. First, inter-user interference becomes challenging to mitigate as more CUs are served simultaneously. Second, sharing the fixed power budget among more users makes it more difficult to satisfy the QoS constraints of all CUs simultaneously.  Notably, SIM-assisted schemes exhibit a significantly slower degradation in secrecy performance compared with their non-SIM counterparts, demonstrating the effectiveness of SIM.}

{Fig.~\ref{fig:impact distance} shows the impact of the maximum distance between the BS and the CUs/Eve on the ASR. In this evaluation, the difference $D_{\max} - D_{\min}$ is fixed at $25$ m.} As the service region moves farther from the BS, the ASR decreases due to increased path loss. Nevertheless, SIM-enabled beamforming maintains consistently higher secrecy rates across all distance ranges, highlighting its robustness against propagation attenuation.

\begin{figure}[t]
    \centering \includegraphics[width=0.85\columnwidth]{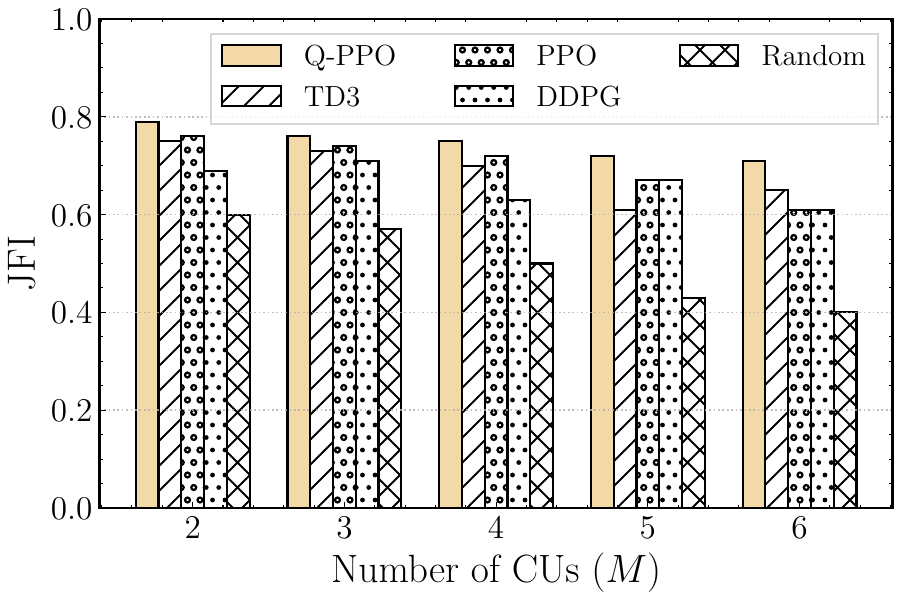}
    \caption{JFI comparison versus the number of CUs, $M$.}
    \label{fig:JFI}
\end{figure}

Fig.~\ref{fig:JFI} compares the fairness index achieved by different algorithms. Since JFI is limited by the rate of the user with the poorest channel condition, considering the minimum rate constraint for each user ensures fairness, with the index consistently exceeding $0.6$. Notably, the proposed Q-PPO algorithm outperforms all other approaches, achieving a fairness level greater than $0.7$ even as the number of users increases.

\begin{figure}[t]
    \centering \includegraphics[width=0.95\columnwidth]{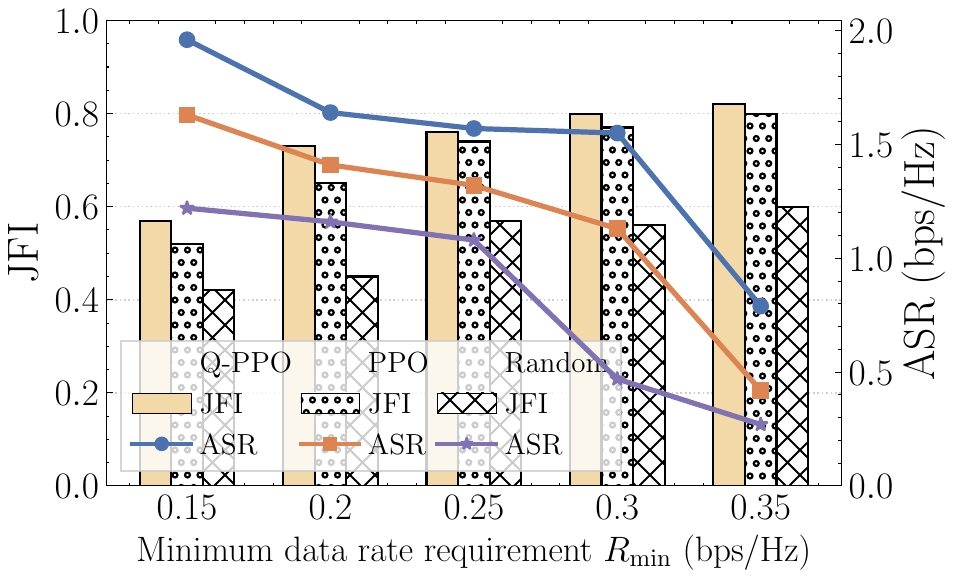}
    \caption{Evolution of ASR and JFI with varying $R_{\text{min}}$.}
    \label{fig:tradeJFI}
\end{figure}

Finally, we investigate the tradeoff between ASR maximization and user fairness by evaluating the impact of the minimum data rate requirement $R_{\min}$ on both the achieved ASR and JFI for all considered schemes. As shown in Fig.~\ref{fig:tradeJFI}, increasing $R_{\min}$ leads to a gradual reduction in ASR while simultaneously improving fairness, highlighting the inherent tradeoff between secrecy performance and fair resource allocation among users. This behavior can be explained by the stricter QoS requirements imposed at larger $R_{\min}$ values. In particular, the system must allocate more transmit power and SIM phase shift resources to users experiencing weaker channel conditions in order to satisfy the minimum rate constraints, which reduces the overall secrecy-rate efficiency. At the same time, this resource redistribution improves fairness among users, as reflected by the increasing JFI values across all schemes. Specifically, the JFI achieved by Q-PPO, PPO, and Random increases from approximately $0.57$, $0.52$, and $0.42$ to $0.82$, $0.80$, and $0.61$, respectively, as $R_{\min}$ increases from $0.15$ to $0.35$ bps/Hz. Despite this tradeoff, the proposed Q-PPO framework consistently achieves the highest ASR while also maintaining the best fairness performance over all considered $R_{\min}$ values. These results demonstrate that the proposed framework achieves a more effective balance between secrecy-rate maximization and user fairness compared with the benchmark schemes.

\section{Conclusions}\label{sec:conclusion}
This work investigated SIMs for enhancing physical-layer security in multi-user MISO systems. By enabling multi-stage wave-domain control, SIMs provide substantially higher degrees of freedom than conventional single-layer surfaces, offering an effective mechanism for suppressing multi-user interference and improving secrecy performance in dense wireless environments. To manage the resulting high-dimensional and strongly coupled control problem, we developed a quantum-enhanced DRL framework that departs from traditional optimization and classical DRL approaches. The proposed Q-PPO algorithm employs a hybrid quantum–classical policy model to overcome scalability limitations and enables adaptive SIM configuration under imperfect channel knowledge. Simulation results demonstrate that SIM-assisted transmission significantly improves secrecy performance and fairness, while Q-PPO achieves faster convergence and stronger adaptability than state-of-the-art DRL baselines. These findings indicate that quantum-enhanced DRL is a promising paradigm for controlling large-scale programmable metasurface architectures and for enabling secure next-generation wireless networks.

\bibliographystyle{IEEEtran}
\bibliography{bibo}
\end{document}